\documentclass[aps,preprint,amssymb,prb]{revtex4}

\usepackage{epsfig}

\def\sca{{\lambda}}
\def\is{{\alpha}}
\def\iw{{\beta}}

\def\rv{{\bf r}}
\def\uv{{\bf u}}

\def\fv{{\bf f}}

\def\beq{\begin{equation}}
\def\eeq{\end{equation}}

\def\nf{N_e}

\begin{document}
\renewcommand{\thefootnote}{\fnsymbol{footnote}} 
\renewcommand{\theequation}{\arabic{section}.\arabic{equation}}

\title{Density functional theory for strongly-interacting electrons: Perspectives for Physics and Chemistry}

\author{Paola Gori-Giorgi$^a$ and Michael Seidl$^b$}


\affiliation{$^a$ Department of Theoretical Chemistry and Amsterdam Center for Multiscale Modeling, FEW, Vrije Universiteit, De Boelelaan 1083, 1081HV Amsterdam, The Netherlands \\
$^b$ Institute of Theoretical Physics, University of Regensburg, D-93040 Regensburg, Germany}

\date{\today}

\begin{abstract}
\noindent
Improving the accuracy and thus broadening the applicability of electronic density functional theory (DFT) is crucial to many research areas, from material science, to theoretical chemistry, biophysics and biochemistry.
In the last three years, the mathematical structure of the strong-interaction limit of density functional theory has been uncovered, and exact information on this limit has started to become available. The aim of this paper is to give a perspective on how this new piece of exact information can be used to treat situations that are problematic for standard Kohn-Sham  DFT. 
One way to use the strong-interaction limit, more relevant for solid-state physical devices, is to define a new framework to do practical, non-conventional, DFT calculations in which a strong-interacting reference system is used instead of the traditional non-interacting one of Kohn and Sham. Another way to proceed, more related to chemical applications, is to include the exact treatment of the strong-interaction limit into approximate exchange-correlation energy density functionals in order to describe difficult situations such as the breaking of the chemical bond.

\end{abstract}

\maketitle

\section{Introduction}
\label{intro}
Density functional theory (DFT),\cite{Koh-RMP-99} in its Kohn-Sham (KS) formulation,\cite{KohSha-PR-65} has been a real breakthrough for electronic structure calculations, allowing to treat systems much larger than those accessible to wavefunction methods. KS DFT, together with its extension to time-dependent (TD) phenomena (TDDFT),\cite{RunGro-PRL-84} made possible the theoretical study of an incredible huge number of chemical, physical, and biological processes. 

The key idea of KS DFT is an exact mapping \cite{KohSha-PR-65} between the physical, interacting, many-electron system and a model system of non-interacting fermions with the same density. Only one term, the so called exchange-correlation (xc) energy functional (containing all the complicated many-body effects) needs to be approximated. Although in principle this functional is unique (or ``universal''), a large number of approximations have been developed in the last twenty years, both by chemists and physicists, often 
targeting different systems, different properties, and different phenomena.  In a way, the emergence of such a ``functional zoology'' simply reflects the intrinsic difficulty of building a single general approximation able to recognize and capture, for each given system or process, the many-body effects relevant for its description. 
 
Despite the large number of available approximate functionals and of their successful 
applications, there are still important cases in which KS DFT can fail, which is why the quest for better xc 
functionals continues to be a very active research field (for recent reviews see, e.g., Refs.~\onlinecite{Mat-SCI-02,PerRuzTaoStaScuCso-JCP-05,BecJoh-JCP-07,ZhaSchTru-JCTC-06,CohMorYan-SCI-08}). For example, present-day KS DFT encounters problems in the treatment of near-degeneracy effects (rearrangement of electrons within partially filled levels, important for describing bond dissociation but also equilibrium geometries, particularly for systems with $d$ and $f$ unsaturated shells), in the description of van der Waals long-range interactions (relevant, for example, for biomolecules and layered materials), and of localization effects due to strong electronic correlations (as those occurring in Mott insulators and in low-density nanodevices, but also occurring in bond dissociation). These problems can hamper more or less severely (and sometimes in an unpredictable way) a given calculation, depending on their relative importance with respect to other effects that are better captured by the available approximate functionals.

This work primarily aims at describing a different approach to some of the unsolved problems of present-day DFT, focussing on the treatment of systems with strong spatial correlations. The key idea is to 
recognize that the non-interacting Kohn-Sham reference system is not always the best choice. The main idea 
of Kohn and Sham, which can be summarized as ``Let's solve a model system having the same density of the 
physical one and approximate the remaining missing energy with a density functional'', can be rigorously 
generalized to model systems different from the non-interacting one of Kohn and Sham.\cite{Sav-INC-96} This freedom can be 
used to choose model systems that are able to capture some of the relevant effects (for example near-degeneracy or strong correlations), whose computational cost is still low, and for which it is easier to design approximate density functionals that recover the missing energy.  
For example, in recent years this strategy has been used to address the problems of near-degeneracy effects 
and van der Waals interactions by using a model system with a weak long-range-only interaction (and having the same density of the physical system, as in KS theory). The preliminary results are so far very successful,\cite{LeiStoWerSav-CPL-97,PolSavLeiSto-JCP-02,AngGerSavTou-PRA-05,GolWerSto-PCCP-05,GolWerStoLeiGorSav-CP-06,FroTouJen-JCP-07,TouGerJanSavAng-PRL-09,JanHenScu-JCP-09} as proved by the growing number of research groups that are now working on the practical implementation of this ``short-range DFT - long-range wavefunction'' (srDFT-lrWF) method.\cite{GolWerSto-PCCP-05,GolWerStoLeiGorSav-CP-06,LivBae-PCCP-07,FroTouJen-JCP-07,GolStoThiSch-PRA-07,GolLeiManMitWerSto-PCCP-08,TouGerJanSavAng-PRL-09,JanHenScu-JCP-09,FroCimJen-PRA-10,PaiJanHenScuGruKre-JCP-10,ZhuTouSavAng-JCP-10}

Strong correlations, however, remain a big challenge for DFT, and in many cases are also beyond the reach of the srDFT-lrWF method.  By ``strong electronic correlation'' we mean here the study of systems in which the 
electron-electron interaction largely dominates over the kinetic energy, creating strong spatial correlations. In such cases, it may happen that we need  very many (billions) of Slater determinants for a proper description of the relevant physics, with all the natural occupation numbers becoming very small. 
For these situations both the non-interacting KS system and the weak-interacting hamiltonian of the srDFT-lrWF method are not the best starting point: they are not able to capture the physics of the system under study so that trying to describe the missing energy with an approximate density functional is often a daunting task (or, alternatively, the srDTF-lrWF method becomes as expensive as solving the Schr\"oedinger equation for the physical system).

\begin{figure}
  \begin{center}
   \includegraphics[width=13cm]{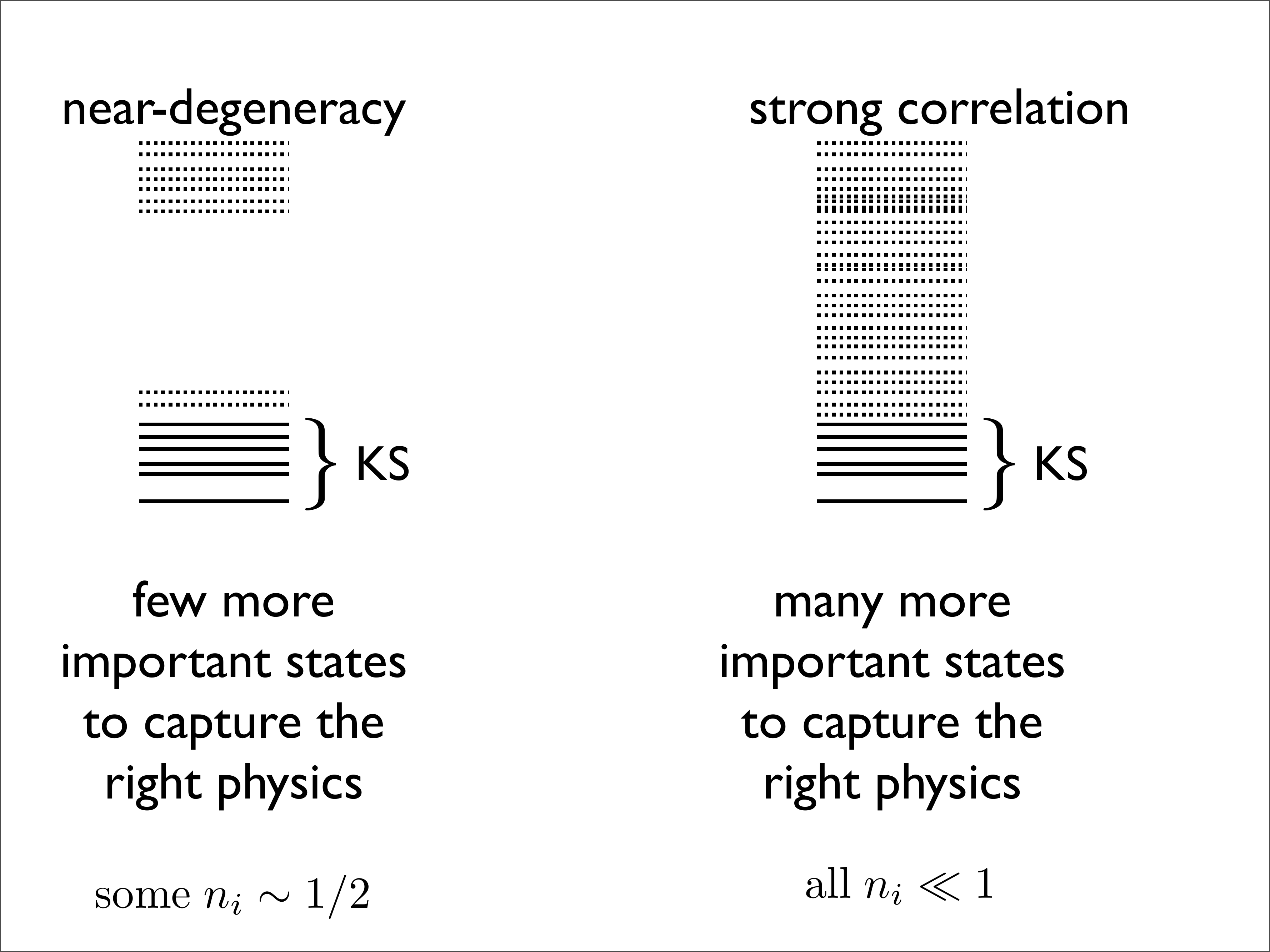}
   \caption{Schematic illustration of the difference between near-degeneracy effects, in which few more important states with respect to the Kohn-Sham occupied levels are needed in order to capture the right physics, and strongly-correlated systems, which need billions of Slater determinants. The first case is usually characterized by the presence of natural occupation numbers $n_i$ close to $1/2$, while the second case often corresponds to natural occupations that are all very small. In this figure levels drawn with a solid line represent the occupied KS states, and dotted levels the empty ones. On the left, we have a typical near-degenerate system, in which few empty states strongly couple to the ground state: including them would be enough to describe the right physics of the system, although for an accurate energy many more states would be needed. On the right we have a strongly correlated system in which billion of states are strongly coupled to the ground state.}
\label{fig_livelli}
  \end{center}
\end{figure}

In order to ``visualize'' this concept, Fig.~\ref{fig_livelli} schematically represents the difference between near-degeneracy effects, characterized by the presence of few more important states with respect to the occupied KS orbitals (that can be captured with a weak-interacting hamiltonian, like the one used in the srDFT-lrWF method), and strong correlations, where very many (billions) of Slater determinants are needed for a proper description of the relevant physics (notice that here we are not talking about getting the energy with high accuracy, but only about describing the right physics: once we have a model hamiltonian which is able to do that, the idea is, as in KS theory, to correct the energy with a density functional).
In this figure levels drawn with a solid line represent the occupied KS states (labeled with ``KS''), and dotted levels the empty ones. On the left, we have a typical near-degenerate system, in which few empty states strongly couple to the ground state: including them would be enough to describe the right physics of the system, although for an accurate energy many more states would be needed. On the right we have a strongly correlated system in which billion of states are strongly coupled to the ground state. From the point of view of the exact first-order density matrix, the first case corresponds to having some natural occupation numbers $n_i$ close to $1/2$ (if we consider natural spin orbitals with $0\le n_i\le 1$), while the second case corresponds to having all $n_i\ll 1$.
Of course this simple, schematic, picture may be very different if we use a spin-unrestricted formalism to define the KS system (see also Sec.~\ref{sec_spinDFT}), instead of a restricted one, as mostly used throughout this paper.

Prototype systems displaying near-degeneracy effects are the Be isoelectronic series (where the 2$s$ and the 2$p$ KS levels become more and more degenerate as the atomic number $Z$ increases), and the H$_2$ molecule along its dissociation curve, where the $\sigma_g$ and $\sigma_u$ KS energies get closer and closer as the molecule is stretched. These two simple examples are paradigmatic of many situations occurring in the study of chemical and physical problems, from heavy elements to the stretching of the chemical bond in general. 
A simple example of strong electronic correlation are low density nanodevices such as quantum dots. As the electronic density is lowered, spatial correlations between the electrons become stronger and stronger, and, as shown in Refs.~\onlinecite{CioPer-JCP-00,CioBuc-JCP-06} for a simple model consisting of two electrons in an harmonic potential, all the natural occupation numbers become very small, indicating the presence of an infinite number of important states. In real systems studied in experiments, in which low-density electrons are confined at the interfaces of semiconductor heterostructures, this phenomenon leads, for example, to intriguing patterns in the addition energy spectra,\cite{ZhiAshPfeWes-PRL-97} which are suggestive of strong spatial correlations and have never been fully explained.

Of course, in general there are very many different physical situations which need a huge number of Slater determinants to be described, and many corresponding ansatz wavefunctions, models or methods that can do that, each one being able to capture different physical phenomena. Typical examples are the density-matrix renormalization group (DMRG) method, the Laughlin wavefunction, the unrestricted Hartree-Fock plus symmetry restoration wavefunction, and dynamical mean field theory.

The main object of this paper is to review and discuss the perspectives of a new way to deal with the case of strong spatial correlations in a DFT framework. For a given $N$-electron system with density $\rho(\rv)$, we construct, in a mathematical rigorous way, a model system consisting of $N$ electrons having the same density $\rho(\rv)$ and maximum possible correlation between the $N$ electronic positions. We call this model system the ``strictly correlated electron'' (SCE) model, and we use it as a complementary alternative to the KS ansatz for DFT. We also propose simple approximate density functionals to recover the difference between the energies of the physical system and of the SCE model,  following the same ideas used in KS DFT. The SCE model is able to capture the infinitely many Slater determinants needed to describe strong spatial correlations, and, as we shall see in the next sections, it is the natural counterpart of the KS ansatz. It also provides a rigorous lower bound for the exact exchange-correlation functional of KS DFT, simply because the electrons cannot be more correlated than the SCE state in a given one electron density $\rho(\rv)$.

The paper is organized as follows. After reviewing the basics of DFT  in Sec.~\ref{basicHK},
in order to emphasize the analogies and the differences between the usual KS DFT and the ``SCE DFT'', we parallel, throughout Secs.~\ref{sec_NIE-SCE}-\ref{sec_exactTsandVeeSCE}, the two approaches. Thus, Secs.~\ref{sec_NIE-SCE}-\ref{sec_exactTsandVeeSCE} contain a KS part, which quickly reviews the main formalism pertinent to the KS ansatz, and a SCE part, which explains how the same concepts can be generalized using the SCE model as a reference system. In Sec.~\ref{sec_calcSCE} we report first applications of the SCE-DFT method to few-electron quantum dots at low density. Although, as previously mentioned, bond dissociation can be viewed as a near-degeneracy effect (which can be described by the weak interacting hamiltonian of the srDFT-lrWF method or, e.g., by density matrix functional theory\cite{GriPerBae-JCP-05,RohPerGriBae-JCP-08} or by a mixture of Hartree-Fock and Hartree-Fock-Bogoliubov methods\cite{TsuScu-JCP-09}), it is also characterized by strong spatial correlations between the electrons involved in the stretched bond, whose physics can be captured by the SCE limit. In
Section~\ref{sec_SCEchem}, thus, we discuss possible ways to include the exact information contained in the SCE limit into functionals useful for chemical applications, with emphasis on bond dissociation. The last Sec.~\ref{sec_conc} is devoted to conclusions and perspectives. 

\section{The Hohenberg-Kohn functional and its basic properties}
\label{basicHK}
We begin by defining the problem and reviewing the basic properties of the Hohenberg-Kohn functional.

We generally consider here systems of $N$ interacting electrons, bound by a given external potential
$v(\rv)$ in $D$-dimensional space ($\rv\in{\sf R}^D$). The corresponding Hamiltonian,
\beq
\hat{H}_\is[v]=\hat{T}+\is\hat{V}_{\rm ee}+\sum_{i=1}^Nv(\rv_i),
\label{eq_Ham}
\eeq
with the universal operators of the kinetic energy,
\beq
\hat{T}=-\frac{\hbar^2}{2m}\sum_{i=1}^N\frac{\partial^2}{\partial\rv_i^2},
\label{eq_TOp}
\eeq
and the Coulomb repulsion between the electrons,
\beq
\hat{V}_{\rm ee}=\frac{e^2}2\sum_{i,j=1}^N\frac{1-\delta_{ij}}{|\rv_i-\rv_j|},
\label{eq_VeeOp}
\eeq
has four independent parameters: the particle number $N$, the spatial dimension $D$, the
${\sf R}^D\to{\sf R}$ function $v=v(\rv)$ of the external potential, and a tunable dimensionless
interaction strength $\is\ge0$ (which will be set to its realistic value $\is=1$ at the end). Unlike
$\is$ and $v$, the parameters $N$ and $D$ will not be indicated explicitly in our notation.

Due to the Ritz principle, the ground-state energy of $\hat{H}_\is[v]$ is given by
\beq
E_\is[v]=\min_{\Psi\to N}\langle\Psi|\hat{H}_\is[v]|\Psi\rangle,
\label{eq_Egs}
\eeq
where the condition $\Psi\to N$ addresses all (normalized) spin-$\frac{1}{2}$ fermionic wave
functions in $D$-dimensional space,
\beq
\Psi=\Psi(\rv_1,...,\rv_N;\sigma_1,...,\sigma_N),
\eeq
with $\rv_i\in{\sf R}^D$ and spin variables $\sigma_i$.
A considerably simpler function is the particle density,
\beq
\rho(\rv)=N\sum_{\sigma_1,...,\sigma_N}\int d^Dr_2...d^Dr_N
\Big|\Psi(\rv,\rv_2,...,\rv_N;\sigma_1,...,\sigma_N)\Big|^2,
\label{rhoDef}
\eeq
which is normalized according to $\int d^Dr\rho(\rv)=N$. In terms of this function as the variable,
the universal Hohenberg-Kohn (HK) functional of DFT is defined as\cite{Lev-PNAS-79,LevPer-INC-85}
\beq
F_\is[\rho]=\min_{\Psi\to\rho}\langle\Psi|\hat{T}+\is\hat{V}_{\rm ee}|\Psi\rangle\;\ge\;0
\label{eq_Fdef}
\eeq
where the condition $\Psi\to\rho$ now addresses only those fermionic $N$-electron wave functions $\Psi$
that are, via Eq.~(\ref{rhoDef}), associated with the same given particle density
$\rho=\rho(\rv)$. Here, ``universal'' means that $F_\is[\rho]$ does not depend on the parameter
$v=v(\rv)$. [It does, however, depend on the spatial dimension $D$ and
on the particle number $N=\int d^Dr\rho(\rv)$.] If the functional $F_\is[\rho]$
was known explicitly in terms of the density $\rho$, the ground-state energy of Eq.~(\ref{eq_Egs})
could be obtained by a considerably simpler minimization procedure,
\beq
E_\is[v]=\min_{\rho\to N}\Big\{F_\is[\rho]+\int d^Dr\rho(\rv)v(\rv)\Big\}
\label{eq_HK}
\eeq
where the condition $\rho\to N$ now addresses all (non-negative) density functions
$\rho(\rv)$ that are normalized to the same given particle number $N$. Eq.~(\ref{eq_HK}) is called
the (second part of the) HK theorem [the first part being the statement that the external potential
$v(\rv)$ in the Hamiltonian of Eq.~(\ref{eq_Ham}) is unambiguously fixed by its ground-state density
$\rho(\rv)$].

Introducing a Lagrangian multiplier $\mu$ to account for the condition $\rho\to N$ (and writing
$F_{\is=1}[\rho]\equiv F[\rho]$), we obtain from Eq.~(\ref{eq_HK}) the Euler equation
\beq
\frac{\delta F[\rho]}{\delta\rho(\rv)}+v(\rv)=\mu,
\label{eq_Euler}
\eeq
to be solved for the wanted density function $\rho(\rv)$. Since $F[\rho]$ is not known
explicitly in terms of the density $\rho$, the crucial problem of DFT is to find approximate ways
of treating $F[\rho]$ and its functional derivative $\delta F[\rho]/\delta\rho(\rv)$.

Clearly, the complexity of the many-body problem is hidden in the HK functional $F_\is[\rho]$.
An equivalent functional is
\beq
\widetilde{F}_\beta[\rho]=\min_{\Psi\to\rho}\langle\Psi|\beta\hat{T}+\hat{V}_{\rm ee}|\Psi\rangle
=\beta F_{1/\beta}[\rho].
\label{eq_FTdef}
\eeq
Since a minimizing wave function here at the same time minimizes Eq.~(\ref{eq_Fdef}) for the
interaction strength $\is=1/\beta$, the parameter $\beta$ may be dubbed the ``interaction 
weakness''.

For a given density $\rho$ and interaction strength $\is$ in Eq.~(\ref{eq_Fdef}), let $\Psi_\is[\rho]$
be a minimizing wave function. 
With $T_\is[\rho]=\langle\Psi_\is[\rho]|\hat{T}|\Psi_\is[\rho]\rangle$ and 
$V^{(\is)}_{\rm ee}[\rho]=\langle\Psi_\is[\rho]|\hat{V}_{\rm ee}|\Psi_\is[\rho]\rangle$ we have
\beq
F_\is[\rho]=T_\is[\rho]+\is V^{(\is)}_{\rm ee}[\rho].
\label{eq_FTV}
\eeq
We make here the usual assumption that $\Psi_\is[\rho]$ depends smoothly on the parameter $\is$. (This assumption may break down, e.g., for a uniform electron gas at low density going through a ferromagnetic phase transition). Then, $F_\is[\rho]$ is differentiable with respect to $\is$ and, due to the minimum property, Eq.~(\ref{eq_Fdef}), the Hellmann-Feynman theorem implies\cite{Har-PRA-84,LanPer-SSC-75,Yan-JCP-98}
\beq
\frac{\rm d}{{\rm d}\is}F_\is[\rho]=\langle\Psi_\is[\rho]|\hat{V}_{\rm ee}|\Psi_\is[\rho]\rangle.
\label{eq_dFdis}
\eeq
In particular, we can write Eq.~(\ref{eq_FTV}), in terms of the universal functionals\cite{LevPer-PRA-85}
\beq
V^{(\is)}_{\rm ee}[\rho]\equiv\frac{\rm d}{{\rm d}\is}F_\is[\rho],\qquad
T_\is[\rho]\equiv F_\is[\rho]-\is\frac{\rm d}{{\rm d}\is}F_\is[\rho].
\label{eq_VeeTdef}
\eeq
An immediate consequence of Eq.~(\ref{eq_dFdis}) is the coupling-constant integral\cite{Har-PRA-84,LanPer-SSC-75,Yan-JCP-98}
\beq
F_1[\rho]-F_0[\rho]=\int_0^1d\is\,V^{(\is)}_{\rm ee}[\rho].
\label{eq_cciV}
\eeq
In an analogous way, the corresponding formula for the functional $\widetilde{F}_\beta[\rho]$ is obtained, 
\beq
F_1[\rho]-\widetilde{F}_0[\rho]=\int_0^1d\iw\,\widetilde{T}_\iw[\rho]
\label{eq_cciT}
\eeq
(notice that $\widetilde{F}_1[\rho]=F_1[\rho]$).
Here, $\widetilde{T}_\iw[\rho]=\langle\widetilde{\Psi}_\iw[\rho]|\hat{T}|\widetilde{\Psi}_\iw[\rho]\rangle$
where $\widetilde{\Psi}_\iw[\rho]=\Psi_{\is=1/\iw}[\rho]$ is a minimizing wave function in
Eq.~(\ref{eq_FTdef}), $\widetilde{T}_\iw[\rho]=T_{\is=1/\iw}[\rho]$. Substituting $\iw=\is^{-1}$, 
we obtain\cite{LiuBur-JCP-09,GorSeiVig-PRL-09}
\beq
F_1[\rho]-\widetilde{F}_0[\rho]=\int_1^\infty\frac{d\is}{\is^2}\,T_\is[\rho].
\eeq

We define a density $\rho$ to be ground-state-$(\is,v)$-representable if there exists a single-particle external potential
$v_\is[\rho](\rv)$ (whose existence is not always granted\cite{Lie-IJQC-83}) such that $\rho$ is a ground-state density of the Hamiltonian
\beq
\hat{H}_\is[\rho]=\hat{T}+\is\hat{V}_{\rm ee}+\sum_{i=1}^Nv_\is[\rho](\rv_i).
\label{eq_Ham2}
\eeq
In this case, $\Psi_\is[\rho]$ is a ground state of $\hat{H}_\is[\rho]$; the corresponding ground-state energy,
\beq
E_\is[\rho]=F_\is[\rho]+\int d^Dr\,v_\is[\rho](\rv)\rho(\rv),
\eeq
however, can be degenerate.

Similarly, the Hamiltonian
\beq
\hat{\widetilde{H}}_\iw[\rho]=\iw\hat{T}+\hat{V}_{\rm ee}+\sum_{i=1}^N\widetilde{v}_\iw[\rho](\rv_i),\qquad
\widetilde{v}_\iw\equiv\iw v_{\is=1/\iw}
\label{eq_Ham3}
\eeq
has the ground state $\widetilde{\Psi}_\iw[\rho]=\Psi_{1/\iw}[\rho]$ and the ground-state energy
\beq
\widetilde{E}_\iw[\rho]=\widetilde{F}_\iw[\rho]+\int d^Dr\,\widetilde{v}_\iw[\rho](\rv)\,\rho(\rv).
\eeq

\section{Zero and strict Coulomb correlation}
\label{sec_NIE-SCE}

\subsection{Non-interacting electrons (NIE)}

The usual Kohn-Sham system corresponds to the non-interacting limit $\is=0$ of the HK functional $F_\is[\rho]$, 
\beq
F_0[\rho]=\lim_{\iw\to\infty}\frac1{\iw}\widetilde{F}_\iw[\rho]
=\min_{\Psi\to\rho}\langle\Psi|\hat{T}|\Psi\rangle\equiv T_{\rm s}[\rho].
\eeq
Being a ground state of the non-interacting Hamiltonian $\hat{H}_{\is=0}[\rho]$, the minimizing
wave function $\Psi_0[\rho]=\widetilde{\Psi}_\infty[\rho]=\Psi^{\rm NIE}[\rho]$ is, in most cases, a
single Slater determinant of $N$ spin-orbitals $\phi_i(\rv,\sigma)$ which obey the Kohn-Sham
(KS) single-particle Schr\"odinger equations
\beq
\Big\{-\frac{\hbar^2}{2m_{\rm e}}\nabla^2+v_0[\rho](\rv)\Big\}\phi_i(\rv,\sigma)=\epsilon_i\phi_i(\rv,\sigma).
\label{eq_KSequ}
\eeq
Consequently, $T_{\rm s}[\rho]$ is the kinetic energy of $N=\int d^Dr\rho(\rv)$ non-interacting
electrons in a given ground-state density $\rho=\rho(\rv)$. By construction, the KS potential
$v_0[\rho](\rv)$ is such that the orbitals reproduce the given density,
\beq
\sum_{i,\sigma}|\phi_i(\rv,\sigma)|^2=\rho(\rv).
\label{eq_phiKSrho}
\eeq
Implicitly, in terms of these orbitals (rather than explicitly in terms of the density $\rho$ itself),
$T_{\rm s}[\rho]$ is given by
\beq
T_{\rm s}[\rho]=\frac{\hbar^2}{2m_{\rm e}}\sum_{i,\sigma}\int d^Dr|\nabla\phi_i(\rv,\sigma)|^2.
\label{eq_TSimpl}
\eeq

Non-interacting electrons (NIE) have zero Coulomb correlation. For example, $N=2$ such electrons in a
given density $\rho(\rv)$ have opposite spins and occupy the same spatial
orbital $\psi(\rv)=\sqrt{\frac12\rho(\rv)}$ (the situation can become more complicated if the corresponding KS potential has a degenerate ground state, something that rarely happens for $N=2$). When their two positions are measured simultaneously,
the results $\rv_1$ and $\rv_2$ are completely uncorrelated -- when only the partial
result $\rv_1$ is noticed while the result $\rv_2$ is ignored or hidden, its probability distribution
is rigorously independent of the particular value of $\rv_1$.
In this case, the expectation of $\hat{V}_{\rm ee}$ is given by
\beq
V_{\rm ee}^{(0)}[\rho]=e^2\int d^Dr_1\int d^Dr_2\,\frac{|\psi(\rv_1)\psi(\rv_2)|^2}{|\rv_1-\rv_2|}
=\frac12U[\rho]\qquad(N=2),
\label{eq_Vee02}
\eeq
with the explicit density functional of the Hartree energy,
\beq
U[\rho]=\frac{e^2}2\int d^Dr\int d^Dr'\,\frac{\rho(\rv)\rho(\rv')}{|\rv-\rv'|}.
\label{eq_U}
\eeq
If the electrons were repulsive bosons (b), an arbitrary number $N$ of them could occupy the same
orbital $\psi(\rv)$. In this case, Eq.~(\ref{eq_Vee02}) would be generalized to 
$V_{\rm ee}^{(0)}[\rho]=V_{\rm bb}^{(0)}[\rho]$ where
\beq
V_{\rm bb}^{(0)}[\rho]=\frac{N-1}N\,U[\rho]\qquad\mbox{(bosons)}.
\label{eq_Vbb0}
\eeq
For $N\ge3$, however, non-interacting electrons must occupy two or more different orbitals. 
Consequently, their positions can no longer be completely uncorrelated. This effect is sometimes called Pauli correlation, since it is not caused by a true repulsive (Coulomb) force between the electrons,
but merely by the Pauli principle. As a result, the true value of $V_{\rm ee}^{(0)}[\rho]$
is for $N\ge3$ lower than the bosonic value of Eq.~(\ref{eq_Vbb0}),
\beq
V_{\rm ee}^{(0)}[\rho]=U[\rho]+E_{\rm x}[\rho]\le\frac{N-1}N U[\rho].
\label{eq_Vee0}
\eeq
The exchange energy $E_{\rm x}[\rho]<0$ is another implicit density functional,
\beq
E_{\rm x}[\rho]=-\frac{e^2}2\sum_{i,j}\delta_{m_s^i,m_s^j}\int d^Dr\int d^Dr'
\frac{\psi^*_i(\rv)\psi_j(\rv)\psi^*_j(\rv')\psi_i(\rv')}{|\rv-\rv'|},
\label{eq_Ex}
\eeq
with $\phi_i(\rv,\sigma)=\psi_i(\rv)\chi_{m_s^i}(\sigma)$.
In Eq.~(\ref{eq_Vee0}), the equal sign, implying $E_{\rm x}[\rho]=-\frac1N U[\rho]$,
holds for $N\le2$, while $E_{\rm x}[\rho]<-\frac1N U[\rho]$ for $N\ge3$.

\subsection{Strictly correlated electrons (SCE)}
In the case $\is>0$, the Coulomb repulsion between the electrons is turned on in the Hamiltonian
$\hat{H}_\is[\rho]$ of Eq.~(\ref{eq_Ham2}). Now, the ground state $\Psi_{\is}[\rho]$ has, in addition
to Pauli correlation (for $N\ge3$), also true Coulomb correlation which is caused by a
repulsive force which lowers the value of $V_{\rm ee}^{(\is)}[\rho]$ as $\is$ grows.
Here we consider the extreme limit $\alpha\to\infty$ of infinitely strong
repulsion,\cite{SeiPerLev-PRA-99,Sei-PRA-99} which we call the ``strictly correlated electrons'' (SCE) limit,
\beq
\lim_{\is\to\infty}\frac1{\is}F_\is[\rho]=\widetilde{F}_0[\rho]
=\min_{\Psi\to\rho}\langle\Psi|\hat{V}_{\rm ee}|\Psi\rangle\equiv V_{\rm ee}^{\rm SCE}[\rho].
\label{eq_VeeSCEdef}
\eeq
The functional $V_{\rm ee}^{\rm SCE}[\rho]$ is the natural counterpart of the KS non-interacting kinetic energy $T_s[\rho]$ and was first addressed about ten years ago,\cite{SeiPerLev-PRA-99,Sei-PRA-99} but only treated in an approximated way, using physically motivated models.\cite{Sei-PRA-99,SeiPerKur-PRA-00} Only recently $V_{\rm ee}^{\rm SCE}[\rho]$ and the square $|\Psi_\infty[\rho]|^2=|\widetilde{\Psi}_0[\rho]|^2=|\Psi^{\rm SCE}[\rho]|^2$ of the corresponding minimizing wave function have been treated exactly in Ref.~\onlinecite{SeiGorSav-PRA-07}, where the interested reader can find more mathematical details. In the following, we summarize the basics of the SCE solution, describing the physics that is captured by $V_{\rm ee}^{\rm SCE}[\rho]$.

$V_{\rm ee}^{\rm SCE}[\rho]$ corresponds to the lowest possible value of the expectation of the electron-electron repulsion in a given density $\rho(\rv)$. In other words, the functional $V_{\rm ee}^{\rm SCE}[\rho]$ defines a classical problem with a {\em given smooth density}. Thus, in contrast to $\Psi_\is[\rho]$ for finite $\is<\infty$, the limiting wave function $\Psi_\infty[\rho]$ does no longer depend on the spin variables $\sigma_1,...,\sigma_N$, and, since the limit is classical (even if it is an unusual classical problem because of the constraint of the smooth density), we can only determine $|\Psi_\infty[\rho]|^2$, which, in terms of the spatial variables $\rv_1,...,\rv_N$, is no
longer a regular function, but rather a Dirac-type distribution, describing electrons
with strictly correlated positions. In practice, this means that the $N$ results $\rv_i\in{\sf R}^D$ ($i=1,...,N$) of a
simultaneous measurement of all electronic positions in the distribution $|\Psi_\infty[\rho]|^2$ are no longer
independent of each other, but strictly related via $N$ so-called co-motion functions $\fv_i(\rv)$,
\beq
\rv_i=\fv_i(\rv)\qquad\Big(\;i=1,...,N;\quad\fv_1(\rv)\equiv\rv\;\Big).
\label{eq_COMOdef}
\eeq
In other words, the position $\rv_1$ of one electron fixes the positions $\rv_i$ ($i> 1$) of all the
others. The co-motion functions obey the group properties \cite{SeiGorSav-PRA-07}
\beq
\Big\{\fv_1(\fv_n(\rv)),...,\fv_N(\fv_n(\rv))\Big\}=\Big\{\fv_1(\rv),...,\fv_N(\rv)\Big\}\qquad
(n=1,...,N),
\eeq
so that Eq.~(\ref{eq_COMOdef}) does not conflict with the symmetry postulate on a wave function
for identical fermions. Moreover, as the position of one of the electrons determines the positions of all the others, the probability of finding one electron at position $\rv$ in the volume element $d^D r$ must be the same of finding the $i^{\rm th}$ electron at position ${\bf f}_i(\rv)$ in the volume element $d^Df_i(\rv)$. 
This means that all the co-motion functions for a given $N$-electron density
$\rho=\rho(\rv)$ must satisfy the differential equation\cite{SeiGorSav-PRA-07}
\beq
\rho(\fv_i(\rv))d^Df_i(\rv)=\rho(\rv)d^Dr\qquad(i=1,...,N),
\label{eq_SCEequ}
\eeq
whose initial conditions are fixed by making the corresponding $V_{\rm ee}^{\rm SCE}[\rho]$, given by
\beq
V_{\rm ee}^{\rm SCE}[\rho]=\frac{e^2}2\sum_{i,j=1}^N
\int d^Dr\,\frac{\rho(\rv)}N\,\frac{1-\delta_{ij}}{|\fv_i(\rv)-\fv_j(\rv)|},
\label{eq_VeeSCE}
\eeq
minimum.\cite{SeiGorSav-PRA-07}. 
Thus, similarly to the $N$ single-particle
orbitals $\phi_i(\rv,\sigma)$ in the NIE Kohn-Sham state,
the co-motion functions $\fv_i(\rv)$ are fixed by the given density function $\rho=\rho(\rv)$.
\cite{SeiGorSav-PRA-07}

Equation~(\ref{eq_VeeSCE}) should be viewed as the counterpart of Eq.~(\ref{eq_TSimpl}) which, also implicitly, represents the density functional $T_{\rm s}[\rho]\equiv T^{\rm NIE}[\rho]$ for the non-interacting kinetic energy in terms of the orbitals $\phi_i(\rv,\sigma)$. The latter represent the counterpart of the co-motion functions $\fv_i(\rv)$ in Eq.~(\ref{eq_VeeSCE}). The counterpart of Eqs.~(\ref{eq_Vee0}) and (\ref{eq_Ex}) for the functional $V_{\rm ee}^{(0)}[\rho]$, in contrast, is the limit $\is\to\infty$ of $T_\is[\rho]$, which, as we shall see later, must be treated with some care since it
diverges but still yields a finite ``first-order'' correction to the energy functional $V_{\rm ee}^{\rm SCE}[\rho]$.
\begin{figure}
  \begin{center}
   \includegraphics[width=13cm]{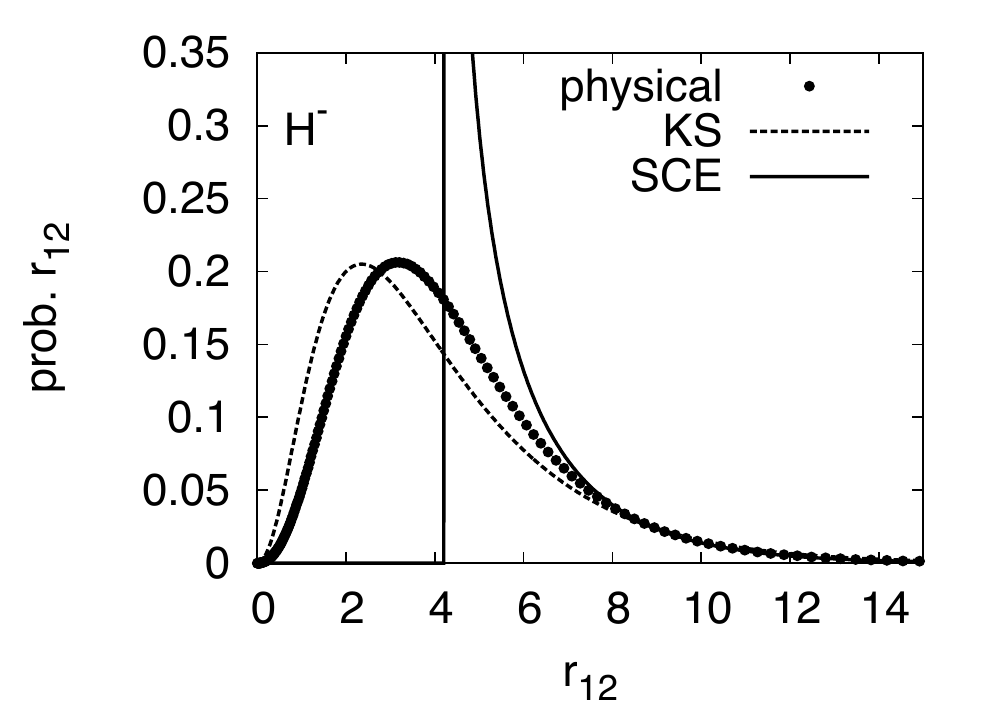}
   \caption{The probability distribution for the electron-electron distance $r_{12}$ for the H$^-$ anion calculated with a very accurate wavefunction for the physical system, with the ``exact'' Kohn-Sham (KS) Slater determinant (built from a very accurate density), and with the strictly correlated electron (SCE) construction. All quantities are in Hartree atomic units.}
\label{fig_intraHminus}
  \end{center}
\end{figure}

The two functionals $T_s[\rho]$ and $V_{\rm ee}^{\rm SCE}[\rho]$ define two different and complementary model systems in which the one-electron density is the same. A simple way to grasp the very different physics captured by the two model systems is to look at the probability density $P(r_{12})$ of finding two electrons at a distance $r_{12}$. As an example, in Fig.~\ref{fig_intraHminus} we report this probability $P(r_{12})$ for the H$^-$ anion calculated using a very accurate wavefunction for the physical system (see Refs.~\onlinecite{FreHuxMor-PRA-84,GorSav-PRA-05} and references therein), using the Kohn-Sham non-interacting Slater determinant (constructed from the same accurate density), and using the SCE construction (see also Ref.~\onlinecite{GorSeiSav-PCCP-08}). The three probabilities  $P(r_{12})$ correspond to three systems having the same one-electron density, that is, the same probability to find one electron at $\rv$ in the volume element $d^3r$. As we see from Fig.~\ref{fig_intraHminus}, the probability distribution for the electron-electron distance is very different: in the KS system there is a higher probability of finding the two electrons close to each other than in the physical system, in which there is Coulomb repulsion. In the SCE state, the two electrons never get closer than a certain distance $r_0\approx 4.2$ a.u., and they avoid each other as much as possible without breaking the constraint of being in the given one-electron density. 

\subsection{Density scaling}

For a given density $\rho=\rho(\rv)$, we consider the usual continuous set of scaled densities $\rho_\sca(\rv)$,
\beq
\rho_\sca(\rv)=\sca^D\rho(\sca\rv)\qquad(\sca>0).
\eeq
The prefactor $\sca^D$ guarantees that $\int d^Dr\,\rho_\sca(\rv)=\int d^Dr\,\rho(\rv)$ for all $\sca>0$.

As the orbitals $\phi_i(\rv,\sigma)$ solve the KS equations Eq.~(\ref{eq_KSequ}) and yield in
Eq.~(\ref{eq_phiKSrho}) the density $\rho(\rv)$, the scaled orbitals $\phi^{(\sca)}_i(\rv,\sigma)=
\sca^{D/2}\phi_i(\sca\rv,\sigma)$ yield the scaled density $\rho_\sca(\rv)$,
\beq
\sum_{i=1}^N\;\sum_\sigma|\phi^{(\sca)}_i(\rv,\sigma)|^2=\rho_\sca(\rv),
\label{eq_phiKSrhosca}
\eeq
and solve the modified KS equations
\beq
\Big\{-\frac{\hbar^2}{2m_{\rm e}}\nabla^2+v_0[\rho_\sca](\rv)\Big\}\phi^{(\sca)}_i(\rv,\sigma)=
\epsilon^{(\sca)}_i\phi^{(\sca)}_i(\rv,\sigma),
\label{eq_KSequsca}
\eeq
where $v_0[\rho_\sca](\rv)=\sca^2v_0[\rho](\sca\rv)$ and $\epsilon^{(\sca)}_i=\sca^2\epsilon_i$.
Therefore, Eq.~(\ref{eq_TSimpl}) implies\cite{Lev-INC-87}    
\beq
T_{\rm s}[\rho_\sca]=\sca^2T_{\rm s}[\rho].
\label{eq_Tssca}
\eeq
For completeness, we note that\cite{Lev-INC-87,LevPer-PRA-85}
\beq
U[\rho_\sca]=\sca U[\rho],\qquad E_{\rm x}[\rho_\sca]=\sca E_{\rm x}[\rho],\qquad
V_{\rm ee}^{(0)}[\rho_\sca]=\sca V_{\rm ee}^{(0)}[\rho].
\label{eq_Usca}
\eeq

Similarly, as the co-motion functions $\fv_i(\rv)$ solve the SCE equation (\ref{eq_SCEequ}) for
the density $\rho(\rv)$, the scaled co-motion functions
\beq
\fv^{(\sca)}_i(\rv)=\frac1{\sca}\fv_i(\sca\rv)
\eeq
solve the corresponding equations for the scaled density $\rho_\sca(\rv)$.
Consequently, Eq.~(\ref{eq_VeeSCE}) implies the scaling behavior
\beq
V_{\rm ee}^{\rm SCE}[\rho_\sca]=\sca V_{\rm ee}^{\rm SCE}[\rho].
\label{eq_VeeSCEsca}
\eeq

We notice that the HK functional has a more involved scaling behavior,\cite{Lev-INC-87}
\beq
F[\rho_\sca]=\sca^2F_{\is}[\rho]\qquad(\is=\sca^{-1}),
\label{eq_Fsca}
\eeq
which is an immediate consequence of Eq.~(\ref{eq_Fdef}) with Eqs.~(\ref{eq_TOp}) and
(\ref{eq_VeeOp}). Thus, for finite $\is$ ($0<\is<\infty$), we could, without
loss of generality, confine ourselves to the case $\is=1$.

\section{Weak and strong Coulomb correlation}
\label{sec_weakandstrong}
Dropping the subscripts $\is$ and the superscript $(\is)$ in Eq.~(\ref{eq_FTV}), we now address
the realistic situation with interaction strength $\is=1$,
\beq
F[\rho]=T[\rho]+V_{\rm ee}[\rho].
\eeq
Here, $F[\rho]=F_{\is=1}[\rho]$, etc. For $N\ge2$, the two contributions on the right-hand side obey the relations
\begin{eqnarray}
T[\rho]=\langle\Psi_{\is=1}[\rho]|\hat{T}|\Psi_{\is=1}[\rho]\rangle
&\ge & T_{\rm s}[\rho]\equiv\min_{\Psi\to\rho}\langle\Psi|\hat{T}|\Psi\rangle\ge 0,\label{eq_Test}\\
V_{\rm ee}[\rho]=\langle\Psi_{\is=1}[\rho]|\hat{V}_{\rm ee}|\Psi_{\is=1}[\rho]\rangle
&\ge & V_{\rm ee}^{\rm SCE}[\rho]\equiv\min_{\Psi\to\rho}\langle\Psi|\hat{V}_{\rm ee}|\Psi\rangle \ge 0.
\label{eq_Veeest}
\end{eqnarray}
(In the trivial case $N=1$, of course, we have $T[\rho]=T_{\rm s}[\rho]>0$ and
$V_{\rm ee}[\rho]=V_{\rm ee}^{\rm SCE}[\rho]=0$.)
These inequalities hold, since the realistic wave function $\Psi_{\is=1}[\rho]$ is significally different
from each one of the two minimizing wave functions on the right-hand side, $\Psi_{\is=0}[\rho]=\Psi^{\rm NIE}[\rho]$
and $\Psi_{\infty}[\rho]=\Psi^{\rm SCE}[\rho]$, respectively. While the latter ones are characterized
completely by $N$ single-particle orbitals $\phi_i(\rv,\sigma)$ or, respectively, by $N$ co-motion
functions $\fv_i(\rv)$, the realistic wave function $\Psi_{\is=1}[\rho]$ is mathematically much more
involved. Describing electrons with finite Coulomb repulsion, it has neither zero nor strict, but rather
some finite Coulomb correlation, a situation which is much harder to describe mathematically. 

The non-interacting kinetic energy $T_{\rm s}[\rho]$ in Eq.~(\ref{eq_Test}) can be considered as the
zero-point kinetic energy resulting (by the uncertainty principle) from the spatial confinement of
non-interacting electrons in the density $\rho=\rho(\rv)$.
For interacting electrons ($\is=1$), this zero-point energy is increased by Coulomb correlation,
since one such electron, due to the repulsion by the other ones, has less effective space available
than a non-interacting one ($\is=0$) within the same given density $\rho=\rho(\rv)$. Consequently,
the resulting difference,
\beq
T_{\rm c}[\rho]=T[\rho]-T_{\rm s}[\rho]>0,
\eeq
is called kinetic energy due to correlation. [We note in passing that, as $\is\to\infty$ grows beyond its
realistic value $\is=1$, this zero-point energy grows indefinitely, see Eq.~(\ref{eq_Tinf}) below.]

On the other hand, increasing Coulomb repulsion ($\is\to\infty$) lowers the expectation of the operator
$\hat{V}_{\rm ee}$ (which is a measure for the average inverse distance $|\rv-\rv'|^{-1}$ between two
electrons
in the state $\Psi_\is[\rho]$). The second inequality, Eq.~(\ref{eq_Veeest}), expresses the fact that 
this lowering is maximum in the limit $\is\to\infty$ of strict correlation, while it is lesser in
realistic systems with $\is=1$ and finite correlation. Therefore, following Ref.~\onlinecite{LiuBur-JCP-09}, the difference
\beq
V_{\rm d}[\rho]=V_{\rm ee}[\rho]-V_{\rm ee}^{\rm SCE}[\rho]>0
\label{eq_defVd}
\eeq
is called decorrelation energy.\cite{GorSeiVig-PRL-09}

Combining the fundamental scaling law of Eq.~(\ref{eq_Fsca}) with the expressions in Eq.~(\ref{eq_VeeTdef}),
one finds the individual scaling properties of the functionals $T[\rho]$ and $V_{\rm ee}[\rho]$,
\beq
\left.\begin{array}{rcl}
T[\rho_\sca]&=&\sca^2T_\is[\rho],\\
V_{\rm ee}[\rho_\sca]&=&\sca V_{\rm ee}^{(\is)}[\rho]\end{array}\right\}
\qquad(\is=\sca^{-1}),
\label{eq_VeeTsca}
\eeq
in contrast to Eqs.~(\ref{eq_Tssca}) and (\ref{eq_VeeSCEsca}).
From section \ref{sec_NIE-SCE}, we know the finite limits
\beq
\lim_{\is\to0}T_{\is}[\rho]=T_{\rm s}[\rho],\qquad
\lim_{\is\to0}V_{\rm ee}^{(\is)}[\rho]=V_{\rm ee}^{(0)}[\rho],\qquad
\lim_{\is\to\infty}V_{\rm ee}^{(\is)}[\rho]=V_{\rm ee}^{\rm SCE}[\rho].
\eeq
In addition, we have the divergent limit \cite{GorVigSei-JCTC-09}
\beq
\is\to\infty:\quad T_\is[\rho]\to T_{\rm ZP}[\rho]\is^{1/2}+O(\is^0),
\label{eq_Tinf}
\eeq
where $T_{\rm ZP}[\rho]$ is the leading coefficient of the expansion describing
zero-point oscillations of strictly correlated electrons about the SCE limit.\cite{GorVigSei-JCTC-09}
Consequently, the high-density limit (HDL) of Eq.~(\ref{eq_VeeTsca}) reads
\beq
\sca\to\infty:\left\{\begin{array}{rcl}
T[\rho_\sca]          &\to& \sca^2T_{\rm s}[\rho]=T_{\rm s}[\rho_\sca],\\
V_{\rm ee}[\rho_\sca] &\to& \sca V_{\rm ee}^{(0)}[\rho]=V_{\rm ee}^{(0)}[\rho_\sca].\end{array}\right.
\label{eq_HDL}
\eeq
In the low-density limit (LDL), in contrast, we have
\beq
\sca\to0:\left\{\begin{array}{rcl}
T[\rho_\sca]          &\to& \sca^{3/2}T_{\rm ZP}[\rho]=T_{\rm ZP}[\rho_\sca],\\
V_{\rm ee}[\rho_\sca] &\to& \sca V_{\rm ee}^{\rm SCE}[\rho]=V_{\rm ee}^{\rm SCE}[\rho_\sca].\end{array}\right.
\label{eq_LDL}
\eeq
Here, we have used Eqs.~(\ref{eq_Tssca}), (\ref{eq_Usca}), (\ref{eq_VeeSCEsca}), and the relation
$T_{\rm ZP}[\rho_\sca]=\sca^{3/2}T_{\rm ZP}[\rho]$ from Ref.~\onlinecite{GorVigSei-JCTC-09}.

Now, we see that the kinetic energy $T[\rho_\sca]$ in the HK functional
\beq
F[\rho_\sca]=T[\rho_\sca]+V_{\rm ee}[\rho_\sca]
\eeq
becomes dominant and approaches its non-interacting value
$T_{\rm s}[\rho_\sca]$ in the HDL ($\sca\to\infty$), while in the LDL ($\sca\to0$), the potential
energy $V_{\rm ee}[\rho_\sca]$ becomes dominant and approaches its strictly correlated limit
$V_{\rm ee}^{\rm SCE}[\rho_\sca]$. Therefore, we call an electron system with given ground-state
density $\rho$ weakly correlated (WCOR), when $T[\rho]\gg V_{\rm ee}[\rho]$ or, more precisely,
\beq
F[\rho]\gtrapprox T[\rho]\gtrapprox T_{\rm s}[\rho]\gg V_{\rm ee}[\rho]
\eeq
and strongly correlated (SCOR), when $V_{\rm ee}[\rho]\gg T[\rho]$ or, more precisely,
\beq
F[\rho]\gtrapprox V_{\rm ee}[\rho]\gtrapprox V_{\rm ee}^{\rm SCE}[\rho]\gg T[\rho].
\eeq

\section{Approximating the HK functional}

\subsection{Exchange-correlation (xc) and kinetic-decorrelation (kd) energies}

When the single-particle orbitals $\phi_i(\rv,\sigma)$ of Eq.~(\ref{eq_TSimpl}) and the co-motion
functions $\fv_i(\rv)$ of Eq.~(\ref{eq_VeeSCE}) can be constructed rigorously for any given density
$\rho=\rho(\rv)$, the functionals $T_{\rm s}[\rho]$ and $V_{\rm ee}^{\rm SCE}[\rho]$ can be treated
exactly. Consequently, there are two natural ways of partitioning the HK functional $F[\rho]$.
The usual one of Kohn and Sham,
\beq
F[\rho]=T_{\rm s}[\rho]+E_{\rm xc}^{\rm H}[\rho],\qquad
E_{\rm xc}^{\rm H}[\rho]\equiv T_{\rm c}[\rho]+V_{\rm ee}[\rho],
\label{eq_Fxc}
\eeq
treats $T_{\rm s}[\rho]$ exactly, and looks for an approximation to
the remaining contribution $E_{\rm xc}^{\rm H}[\rho]$. Since $F[\rho]=F_1[\rho]$ and
$T_{\rm s}[\rho]=F_0[\rho]$, Eq.~(\ref{eq_cciV}) now reads
\beq
E_{\rm xc}^{\rm H}[\rho]=\int_0^1d\is\,V^{(\is)}_{\rm ee}[\rho].
\label{eq_ExcH}
\eeq
The KS DFT scheme works well for weakly and moderately correlated systems (WCOR).
For SCOR systems, where $F[\rho]$ is dominated by $V_{\rm ee}^{\rm SCE}[\rho]$, better results should be obtained by partitioning the HK functional as
\beq
F[\rho]=V_{\rm ee}^{\rm SCE}[\rho]+E_{\rm kd}[\rho],\qquad
E_{\rm kd}[\rho]\equiv T[\rho]+V_{\rm d}[\rho],
\label{eq_Fkd}
\eeq
with $V_{\rm ee}^{\rm SCE}[\rho]$ to be treated exactly and $E_{\rm kd}[\rho]$ to be approximated.
Eq.~(\ref{eq_cciT}) now reads
\beq
E_{\rm kd}[\rho]=\int_0^1d\iw\,\widetilde{T}_\iw[\rho]\equiv\int_1^\infty\frac{d\is}{\is^2}\,T_\is[\rho].
\label{eq_HelFeynforT}
\eeq

The natural counterpart of this so-called kinetic-decorrelation (kd) energy\cite{LiuBur-JCP-09,GorSeiVig-PRL-09}   $E_{\rm kd}[\rho]$
is the xc-Hartree energy $E_{\rm xc}^{\rm H}[\rho]$ of
Eqs.~(\ref{eq_Fxc},\ref{eq_ExcH}). This functional is usually written as
\beq
E_{\rm xc}^{\rm H}[\rho]=E_{\rm xc}[\rho]+U[\rho],
\eeq
with the functional of the exchange-correlation (xc) energy,
\beq
E_{\rm xc}[\rho]
=\underbrace{V_{\rm ee}^{(0)}[\rho]-U[\rho]}_{E_{\rm x}[\rho]}
+\underbrace{V_{\rm ee}[\rho]-V_{\rm ee}^{(0)}[\rho]+T_{\rm c}[\rho]}_{E_{\rm c}[\rho]},
\eeq
where we have introduced the correlation energy $E_{\rm c}[\rho]$. An equivalent representation is
\beq
E_{\rm xc}[\rho]=\Big(T[\rho]-T_{\rm s}[\rho]\Big)+\Big(V_{\rm ee}[\rho]-U[\rho]\Big).
\eeq
Note also that
\beq
E_{\rm kd}[\rho]=T_{\rm s}[\rho]+E_{\rm xc}[\rho]-\Big(V_{\rm ee}^{\rm SCE}[\rho]-U[\rho]\Big).
\label{eq_Ekdxc}
\eeq

\subsection{Local-density approximation (LDA) for $E_{\rm xc}[\rho]$ and $E_{\rm kd}[\rho]$}

A simple approximation to the functional $E_{\rm xc}[\rho]$ or, equivalently,
$E_{\rm xc}^{\rm H}[\rho]=E_{\rm xc}[\rho]+U[\rho]$ is the local-density approximation (LDA),
\beq
E^{\rm LDA}_{\rm xc}[\rho]=\int d^Dr\,\rho(\rv)\,\epsilon^{(D)}_{\rm xc}(r_{\rm s}(\rv)).
\label{eq_ExcLDA}
\eeq
As a function of $\rv$, the dimensionless local density parameter $r_{\rm s}(\rv)$ is given by
\beq
r_{\rm s}(\rv)=\Big(\frac1{\rho(\rv)B_D}\Big)^{1/D}\qquad\Leftrightarrow\qquad
\rho(\rv)=\frac1{B_D\,r_{\rm s}(\rv)^D},
\label{eq_rsDef}
\eeq
where $B_D$ is the volume of a $D$-dimensional ball with radius $a_B=\hbar^2/m_{\rm e}e^2$.
E.g.: $B_3=\frac{4\pi}3a_B^3$, $B_2=\pi a_B^2$. The crucial quantity in Eq.~(\ref{eq_ExcLDA})
is $\epsilon^{(D)}_{\rm xc}(r_{\rm s})$, the xc energy per particle in the $D$-dimensional
uniform electron gas with (uniform) density $\bar{\rho}=(B_Dr_{\rm s}^D)^{-1}$.

The functions $\epsilon^{(D)}_{\rm xc}(r_{\rm s})$ for $D=2,3$ are not known analytically, but
accurate parametrizations of numerical Quantum Monte Carlo (QMC) data are available. In the case $D=2$,
the data and parametrization of Attaccalite {\em et al}.\cite{AttMorGorBac-PRL-02} are nowadays widely used. For $D=3$,  popular parametrizations of the Ceperley and Alder QMC data\cite{CepAld-PRL-80} are the ones of Vosko, Wilk and Nusair\cite{VosWilNus-CJP-80} and of Perdew and Wang.\cite{PerWan-PRB-92} Remarkably, the function $\epsilon^{(3)}_{\rm xc}(r_{\rm s})$ can be
interpolated accurately between its high- ($r_{\rm s}\ll1$) and low-density ($r_{\rm s}\gg1$) limits,
almost without relying on any QMC input at all.\cite{SunPerSei-PRB-10} Finally, for $D=1$ parametrized QMC data of the ground state energy of a uniform electron gas with regularized electron-electron interaction are also available.\cite{CasSorSen-PRB-06}

Given $E^{\rm LDA}_{\rm xc}[\rho]$, a corresponding LDA for $E_{\rm kd}[\rho]$ is readily obtained
from Eq.~(\ref{eq_Ekdxc}),\cite{GorSeiVig-PRL-09}
\beq
E^{\rm LDA}_{\rm kd}[\rho]=\int d^Dr\,\rho(\rv)\,\epsilon^{(D)}_{\rm kd}(r_{\rm s}(\rv)),
\label{eq_EkdLDA}
\eeq
with the kd energy per particle in the $D$-dimensional uniform electron gas,
\beq
\epsilon^{(D)}_{\rm kd}(r_{\rm s})=t^{(D)}_{\rm s}(r_{\rm s})+\epsilon^{(D)}_{\rm xc}(r_{\rm s})
-\frac{a^{(D)}_{\rm M}}{r_{\rm s}}.
\eeq
The non-interacting kinetic energy $t^{(D)}_{\rm s}(r_{\rm s})$ per particle in the
uniform electron gas (in units of $1~{\rm Ha}=e^2/a_B=m_{\rm e}e^4/\hbar^2$)
is known analytically,
\begin{eqnarray}
t^{(2)}_{\rm s}(r_{\rm s})&=&\frac12\frac{(1+\zeta)^2+(1-\zeta)^2}{2r_{\rm s}^2}=\frac{1+\zeta^2}{2r_{\rm s}^2},\\
t^{(3)}_{\rm s}(r_{\rm s})&=&\frac3{10}\Big(\frac{9\pi}4\Big)^{2/3}
                           \frac{(1+\zeta)^{5/3}+(1-\zeta)^{5/3}}{2r_{\rm s}^2},
\end{eqnarray}
and the coefficient $a^{(D)}_{\rm M}$ determines the Madelung energy (in units of $1~{\rm Ha}$),
\beq
a^{(2)}_{\rm M}=-1.1061,\qquad a^{(3)}_{\rm M}=-0.89593.
\eeq
The Madelung energy $\frac{a^{(D)}_{\rm M}}{r_{\rm s}}$ exactly corresponds to the thermodynamic limit (number of particles and volume going to infinity with the particle density kept fixed) of $V_{\rm ee}^{\rm SCE}[\rho]/N$ in a uniform electron gas (with the usual cancellation between the Hartree term, the electron-background and the background-background interaction energies). Thus, as in KS theory, the LDA is uniquely defined as the approximation that makes the method exact in the limit of uniform density.

\subsection{Exact first-order approximation for $E_{\rm xc}[\rho]$ and $E_{\rm kd}[\rho]$}
In KS DFT the exact first-order approximation for $E_{\rm xc}[\rho]$ is the exchange energy of Eq.~(\ref{eq_Ex}), which, as said, is an implicit functional of the density through the KS orbitals.

The ``first-order'' approximation for $E_{\rm kd}[\rho]$ corresponds to zero point (ZP) oscillations around the SCE minimum.\cite{GorVigSei-JCTC-09} The proof that this is indeed the exact first-order correction is rather lengthy and the interested reader can find all the details in Ref.~\onlinecite{GorVigSei-JCTC-09}.

Basically, in the SCE limit the total potential energy of a classical configuration  
\beq
E_{pot}(\rv_1,...,\rv_N)= \sum_{i<j}\frac{e^2}{|\rv_i -\rv_j|}+\sum_i v_{\rm SCE}[\rho](\rv_i)\,,
\eeq
 where $v_{\rm SCE}[\rho](\rv)$ is the external potential associated with the density $\rho$ at zero kinetic energy, is constant on the $D$-dimensional subspace $\Omega_0=\{{\bf f}_1(\rv),\dots,{\bf f}_N(\rv)\}$ of the full $ND$-dimensional configuration space\cite{SeiGorSav-PRA-07} and is expected to have a minimum with respect to variations perpendicular to $\Omega_0$, implying that its Hessian has $D$ eigenvectors with null eigenvalue and $ND-D$ eigenvectors with positive eigenvalue $\omega_\mu^2(\rv) $ at every point on $\Omega_0$.\cite{GorVigSei-JCTC-09} In terms of these eigenvalues, the small $\iw$ and the large $\is$ expansion of $\widetilde{T}_\iw[\rho]$ defined after Eq.~(\ref{eq_cciT}) and $T_\is[\rho]$ of Eq.~(\ref{eq_FTV}) read
\begin{eqnarray}
	\lim_{\beta\to 0} \widetilde{T}_\iw[\rho] &= & \iw^{-1/2}T_{\rm ZP}[\rho]+O(\iw^0) 
	\label{eq_Tsmalliw} \\
	\lim_{\alpha\to\infty}T_\is[\rho] & = & \is^{1/2}T_{\rm ZP}[\rho]+O(\is^0),
	\label{eq_Tlargeis}
\end{eqnarray}
with
\beq
T_{\rm ZP}[\rho]=\frac{1}{2}\int d^D r\,\frac{\rho(\rv)}{N}\,\sum_{\mu=1}^{ND-D}\frac{\omega_\mu(\rv)}{2}.
\eeq
Thus, as anticipated in Sec.~\ref{sec_weakandstrong}, in the strict correlation limit the kinetic energy grows indefinitely. However, both Eqs.~(\ref{eq_Tsmalliw}) and (\ref{eq_Tlargeis}) when inserted in Eq.~(\ref{eq_HelFeynforT}) yield the finite result
\beq
E_{\rm kd}^{\rm ZP}[\rho]=2\,T_{\rm ZP}[\rho]=\int d^D r\,\frac{\rho(\rv)}{N}\,\sum_{\mu=1}^{ND-D}\frac{\omega_\mu(\rv)}{2},
\label{eq_EkdZP}
\eeq
which is the SCE counterpart of the exact exchange energy of Eq.~(\ref{eq_Ex}) for KS theory. The energy $E_{\rm kd}^{\rm ZP}[\rho]$ has a highly non trivial functional dependence on $\rho$, so that its functional derivative is not easily accessible.

\section{Exact treatment of $T_{\rm s}[\rho]$ or $V_{\rm ee}^{\rm SCE}[\rho]$}
\label{sec_exactTsandVeeSCE}

\subsection{The Kohn-Sham approach (exact $T_{\rm s}[\rho]$)}
\subsubsection{Spin-restricted formalism}
With Eq.~(\ref{eq_Fxc}) for the HK functional $F[\rho]$, the Euler equation
Eq.~(\ref{eq_Euler}) reads
\beq
\frac{\delta T_{\rm s}[\rho]}{\delta\rho(\rv)}+\Phi[\rho](\rv)+v_{\rm xc}[\rho](\rv)+v(\rv)=\mu,
\label{eq_Eulerxc}
\eeq
with the electrostatic potential
\beq
\Phi[\rho](\rv)\equiv\frac{\delta U[\rho]}{\delta\rho(\rv)}=e^2\int d^Dr'\frac{\rho(\rv')}{|\rv-\rv'|}
\eeq
of the density $\rho(\rv)$ and the xc potential,
\beq
v_{\rm xc}[\rho](\rv)\equiv\frac{\delta E_{\rm xc}[\rho]}{\delta\rho(\rv)}.
\eeq
When the approximation $E_{\rm xc}^{\rm ap}[\rho]$ used for $E_{\rm xc}[\rho]$ is an explicit
density functional, the corresponding functional derivative $v_{\rm xc}^{\rm ap}[\rho](\rv)=
\delta E_{\rm xc}^{\rm ap}[\rho]/\delta\rho(\rv)$ can be evaluated for any given density
function $\rho(\rv)$. 

By varying the density through variations of the orbitals, Eq.~(\ref{eq_Eulerxc}) for interacting electrons is formally equivalent to the
corresponding equation for a system of non-interacting electrons in the KS effective external
potential
\beq
v_{\rm KS}[\rho](\rv)=\Phi[\rho](\rv)+v_{\rm xc}[\rho](\rv)+v(\rv).
\label{eq_vKS}
\eeq
Thus, the KS orbitals satisfy the equations
\beq
\Big\{-\frac{\hbar^2}{2m_{\rm e}}\nabla^2+v_{\rm KS}[\rho](\rv)\Big\}\phi_i(\rv,\sigma)
=\epsilon^{\rm KS}_i\phi_i(\rv,\sigma),
\label{eq_KSequfinal}
\eeq
which have to be solved self-consistently with Eq.~(\ref{eq_phiKSrho}).

Since the exchange-correlation functional must be approximated in practice, one obtains an approximate ground-state energy for the physical interacting system,
$E^{\rm ap}_0=T_{\rm s}[\rho]+(E_{\rm xc}^{\rm ap}[\rho]+U[\rho])
+\int d^Dr\rho(\rv)\,v(\rv)$.

Employing in Eq.~(\ref{eq_KSequfinal}) the exact quantum-mechanical operator of the kinetic energy,
the functional $T_{\rm s}[\rho]$ is treated exactly here. Consequently, this approach works well in the case
of WCOR systems when $T_{\rm s}[\rho]$ is the dominant contribution to $F[\rho]$.
For SCOR systems, in contrast, we will analyze in the next section a complementary approach based on the exact treatment of $V_{ee}^{\rm SCE}[\rho]$. Before doing so, however, we briefly review the widely used spin-DFT (or unrestricted Kohn-Sham) formalism.

\subsubsection{Spin-unrestricted formalism}
\label{sec_spinDFT}
In practical calculations the spin-DFT version\cite{BarHed-JPC-72} of KS DFT is widely used.
Although the Hoehenberg-Kohn functional only depends on the total density $\rho(\rv)$, in spin DFT one introduces the functional $T_s[\rho_\uparrow,\rho_\downarrow]$,
\beq
T_s[\rho_\uparrow,\rho_\downarrow]=\min_{\Psi\to \rho_\uparrow,\rho_\downarrow}\langle\Psi|\hat{T}|\Psi\rangle,
\eeq
which corresponds to the kinetic energy of a non-interacting system having given spin densities $\rho_\uparrow(\rv)$ and $\rho_\downarrow(\rv)$, with
\beq
\rho_{\sigma}(\rv)=N\sum_{\sigma_2,...,\sigma_N}\int d^Dr_2...d^Dr_N
\Big|\Psi(\rv,\rv_2,...,\rv_N;\sigma,\sigma_2,...,\sigma_N)\Big|^2,
\label{eq_rhosigma}
\eeq
and $\rho_\uparrow+\rho_\downarrow=\rho$. The functional $T_s[\rho_\uparrow,\rho_\downarrow]$ can be used to decompose the HK functional as
\beq
F[\rho]=T_s[\rho_\uparrow,\rho_\downarrow]+U[\rho]+E_{\rm xc}[\rho_\uparrow,\rho_\downarrow]+\int d^D r\, v({\rv})\,\rho(\rv),
\label{eq_UKS}
\eeq 
where $E_{\rm xc}[\rho_\uparrow,\rho_\downarrow]$ is defined as the correction needed to make Eq.~(\ref{eq_UKS}) exact. The idea is to have a non-interacting system with the same spin densities of the true, interacting, one. This constraint defines two effective potentials $v_{\rm KS,\uparrow}[\rho](\rv)$ and $v_{\rm KS,\downarrow}[\rho](\rv)$, and two sets of orbitals such that $\sum_i|\phi_{i,\sigma}(\rv)|^2=\rho_{\sigma}(\rv)$. 

Notice that we have (for the exact functionals evaluated at the exact density and spin densities) $T_s[\rho_\uparrow,\rho_\downarrow]\ge T_s[\rho]$, $E_{\rm xc}[\rho_\uparrow,\rho_\downarrow]\le E_{\rm xc}[\rho]$, and $T_s[\rho_\uparrow,\rho_\downarrow]+E_{\rm xc}[\rho_\uparrow,\rho_\downarrow]=T_s[\rho]+E_{\rm xc}[\rho]$. Using the spin-unrestricted KS reference system instead of the restricted one allows to mimic some correlation effects, similarly to the spin-unrestricted Hartree Fock method.

\subsection{The SCE approach (exact $V_{\rm ee}^{\rm SCE}[\rho]$)}
\label{subsec_SCEapproach}

The non-interacting functionals $T_{\rm s}[\rho]$ and $T_s[\rho_\uparrow,\rho_\downarrow]$ require 
a self-consistent procedure for their calculation. This is because the density (or the spin densities) is determined by the KS orbitals by the simple equation $\sum_i|\phi_{i}(\rv)|^2=\rho(\rv)$, while determining the orbitals from the density requires a highly non-trivial procedure (for which very many different numerical techniques have been proposed in the last years, e.g.,\cite{vanBae-PRA-94,ZhaMorPar-PRA-94,ColSav-JCP-99}). 

The construction of the complementary functional $V_{\rm ee}^{\rm SCE}[\rho]$ for strictly correlated electrons for a given density $\rho(\rv)$ can be simpler, because the density determines the co-motion functions ${\bf f}_i(\rv)$ via the differential equations (\ref{eq_SCEequ}). In other words, in the SCE case it is easier to determine the co-motion functions from the density than to determine the density from the co-motion functions. In particular, $V_{\rm ee}^{\rm SCE}[\rho]$ has been directly constructed for spherically symmetric densities,\cite{SeiGorSav-PRA-07} while algorithms to solve the SCE equations in the general case are under study: a very promising way to proceed is to exploit the similarity between the SCE problem and mass transportation theory.\cite{ButDepGor-PRA-10}

The problem of calculating $V_{\rm ee}^{\rm SCE}[\rho]$ can be reformulated as\cite{SeiGorSav-PRA-07} 
\beq
V_{\rm ee}^{\rm SCE}[\rho]=\min_{\psi\to\rho} \int |\psi(\rv_1,\rv_2,\dots,\rv_N)|^2 \sum_{j>j}\frac{1}{|\rv_i-\rv_j|},
\eeq
where $|\psi|^2$ is the spatial part of the many-electron wavefunction. As said, in fact, in the SCE case, the electrons are strongly distinguished by their relative positions, so that the spin state (or more generally, the statistics) does not play a role.\cite{SeiGorSav-PRA-07} The functional $V_{\rm ee}^{\rm SCE}[\rho]$ is thus the same as the spin unrestricted functional $V_{\rm ee}^{\rm SCE}[\rho_\uparrow,\rho_\downarrow]$ (of course with $\rho_\uparrow+\rho_\downarrow=\rho$). This means that also the exact kinetic and decorrelation functional $E_{\rm kd}[\rho]$ is the same in the spin restricted and spin unrestricted formalism. However, when we deal with approximations for $E_{\rm kd}[\rho]$ this might not be true. In Sec.~\ref{sec_QDN3}, we will compare the results for a quantum dot with three electrons obtained by using the local spin density functional $E_{\rm kd}^{\rm LSD}[\rho_\uparrow,\rho_\downarrow]$ with those from the LDA functional.

Since the co-motion functions can be constructed from the density, in the SCE approach we can obtain the many-electron energy by directly minimizing the
expression $F[\rho]+\int d^Drv(\rv)\rho(\rv)$ with respect to the density function
$\rho(\rv)$, according to Eq.~(\ref{eq_HK}). To this end, the HK functional $F[\rho]$ must be
partitioned as in Eq.~(\ref{eq_Fkd}) where an approximation $E_{\rm kd}^{\rm ap}[\rho]$ is
required for the functional $E_{\rm kd}[\rho]$,
\beq
E^{\rm ap}[v]=\min_{\rho\to N}\Big\{V_{\rm ee}^{\rm SCE}[\rho]+E_{\rm kd}^{\rm ap}[\rho]+
\int d^Drv(\rv)\rho(\rv)\Big\}.
\label{eq_ESCEmin}
\eeq
Unlike the KS equations, this approach should be particularly suitable for SCOR systems
for which the HK functional is dominated by $V_{\rm ee}^{\rm SCE}[\rho]$. In such cases, the density is dominated by strong spatial correlations rather than by the quantum mechanical shells.
In practical calculations, the minimization of Eq.~(\ref{eq_ESCEmin}) can be carried out by expanding the density on a suitable basis set or by using a grid. A simple example of such a calculation is reported in the next Sec.~\ref{sec_QDN2}. 

Another equation that the minimizing density must satisfy can be obtained by varying the energy with respect to $\rho(\rv)$:
\beq
\frac{\delta E[v]}{\delta \rho(\rv)}=\frac{\delta V_{\rm ee}^{\rm SCE}[\rho]}{\delta \rho(\rv)}+\frac{\delta E_{\rm  kd}[\rho]}{\delta \rho(\rv)}+v(\rv)=\mu,
\label{eq_SCEDFTvar}
\eeq
where $\mu$ is the chemical potential.
Although the functional $V_{\rm ee}^{\rm SCE}[\rho]$ depends on the density in a rather complicated way via the co-motion functions [see Eq.~(\ref{eq_VeeSCE})], its functional derivative $v_{\rm SCE}[\rho](\rv)\equiv -\frac{\delta V_{\rm ee}^{\rm SCE}[\rho]}{\delta \rho(\rv)}$ satisfies the classical equilibrium equation\cite{SeiGorSav-PRA-07}
\beq
\nabla v_{\rm SCE}[\rho](\rv)=\sum_{i= 2}^N \frac{\rv-\fv_i(\rv)}{|\rv-\fv_i(\rv
)|^3},
\label{eq_vSCEeq}
\eeq
which has  a very simple physical meaning: the potential $v_{\rm SCE}[\rho](\rv)$ must compensate the net
force acting on the electron in $\rv$, resulting from the repulsion
of the other $N-1$ electrons at positions $\fv_i(\rv)$. The one-body potential $v_{\rm SCE}[\rho](\rv)$ is the counterpart of the KS effective potential of Eq.~(\ref{eq_vKS}) and corresponds to the Lagrange multiplier for the constraint $\Psi\to\rho$ in the minimization of Eq.~(\ref{eq_VeeSCEdef}).
Thus, another possibility to solve the SCE-DFT equations is to look for the density $\rho(\rv)$ that satisfies Eqs.~(\ref{eq_SCEDFTvar}), (\ref{eq_vSCEeq}) and (\ref{eq_VeeSCE}). This last way to proceed, however, raises some questions about the uniqueness of the solution, questions that will be addressed in future work.

\section{SCE-DFT applied to few-electron quantum dots}
\label{sec_calcSCE}
In this Section we report preliminary applications of the SCE-DFT method on simple quantum dots models with few electrons.

Quantum dots are nanodevices in which the motion of electrons is quantized in all three dimensions through the lateral confinement of a high-mobility modulation-doped two-dimensional electron gas in a semiconductor 
heterostructure (for a review, see, e.g., \cite{ReiMan-RMP-02}). Because the confinement of electrons in these ``artificial atoms'' can be varied 
at will, they have become a playground in which the basic physics of interacting electrons can be largely explored and theoretical models can be tested. 
 The number of confined electrons can vary from a few to several hundred, 
with smaller numbers of electrons becoming increasingly 
technologically important in nandevices such as 
the single-electron transistor.

In quantum dots the correlation effects between electrons need to be considered carefully because the external confinement can become much weaker than in real atoms, where the independent electron model with mean-field theories usually gives good results. As the confinement strength is lowered, 
the mutual Coulomb interaction becomes gradually dominant. The physics of this regime can be thus much better captured by SCE-DFT than by traditional KS-DFT. Indeed,
KS DFT has proved useful for studying quantum dots in the weakly correlated regime (e.g., \cite{ReiMan-RMP-02,JiaBarYan-PRB-03,JiaUllYanBar-PRB-04,RasHarPusNie-PRB-04,PitRasProGro-PRB-09}), while the medium and 
strongly-correlated regime, and in particular the cross-over from the Fermi liquid behavior to the Wigner-crystal-like state, has only been accessible to wavefunction methods, e.g., configuration interaction\cite{ReiMan-RMP-02,RonCavBelGol-JCP-06,BluJos-PRB-10} 
(only for very small dots), Quantum Monte Carlo (e.g., \cite{GhoGucUmrUllBar-NP-06,GucGhoUmrBar-PRB-08,ZenGeiRuaUmrCho-PRB-09}) or unrestricted Hartree-Fock plus symmetry restoration.\cite{YanLan-RPP-07} Here we explore with SCE-DFT the regime of weak confinement (strong correlation), where state-of-the-art KS-DFT breaks down.

We thus consider a simple quantum-dot model consisting of $N$ electrons in two dimensions (2D) laterally confined by a parabolic potential:
\beq
\hat{H}=-\frac{\hbar^2}{2 {m^*}}\sum_{i=1}^N\nabla_i^2+\frac{e^2}{\epsilon}\sum_{i=1}^N\sum_{j=i+1}^N\frac{1}{|\rv_i-\rv_j|}+
m^* \frac{\omega^2}{2}\sum_{i=1}^N r_i^2, 
\label{eq_HQD}
\eeq
where $m^*$ is the effective mass and $\epsilon$ the dielectric constant.  

For now we only analyze single dots for which we obtain circularly symmetric densities, $\rho(\rv)=\rho(r)$. In this case, the problem of determining $V_{\rm ee}^{\rm SCE}[\rho]$ can be separated into an angular part and a radial part.\cite{SeiGorSav-PRA-07} The distance $r$ from the center of the dot of one of the electrons can be freely chosen, and it then determines the distances from the center of all the other $N-1$ electrons via radial co-motion functions $f_i(r)$, as well as all the relative angles $\theta_{ij}(r)$ between the electrons.\cite{SeiGorSav-PRA-07} The radial co-motion functions $f_i(r)$ can be constructed as follows.\cite{SeiGorSav-PRA-07} Define an integer index $k$ running for odd $N$ from 1 to $(N-1)/2$, 
and for even $N$ from 1 to $(N-2)/2$. Then
\begin{eqnarray}
 f_{2k}(r)=\left\{
\begin{array}{lr}
 \nf^{-1}(2k-\nf(r)) & r\leq a_{2 k} \\
  \nf^{-1}(\nf(r)-2k) & r> a_{2k}
\end{array}
\right. \nonumber
 \\
 f_{2k+1}(r)=\left\{
\begin{array}{lr}
 \nf^{-1}(\nf(r)+2k) & r\leq a_{N-2 k} \\
  \nf^{-1}(2N-2k-\nf(r)) & r> a_{N-2k},
\end{array}
\right.
\label{eq_fradial}
\end{eqnarray}
where  $a_i=\nf^{-1}(i)$,  
\beq
\nf(r)=\int_0^r 2 \pi \,x \rho(x)\,dx,
\eeq
and $\nf^{-1}(y)$ is the inverse function of $\nf(r)$.
For odd $N$, these equations give all the needed $N-1$ radial co-motion 
functions, while for even $N$ we have to add the last function,
\beq
f_N(r)= \nf^{-1}(N-\nf(r)).
\label{eq_fradialN}
\eeq
The relative angles $\theta_{ij}(r)$ between the electrons can be found by minimizing numerically the electron-electron repulsion energy $\sum_{i>j} [f_i(r)^2+f_j(r)^2-2f_i(r)f_j(r) \cos\theta_{ij}]^{-1/2}$. The radial co-motion functions of Eqs.~(\ref{eq_fradial})-(\ref{eq_fradialN}) satisfy Eq.~(\ref{eq_SCEequ}) for 2D circularly symmetric $\rho$, 
\beq
2 \pi \,f_i(r) \rho(f_i(r))\,|f'_i(r)|\,dr=2 \pi \,r \rho(r)\,dr,
\label{eq_detailedbalancespher}
\eeq 
and, together with the minimizing angles $\theta_{ij}(r)$, yield the minimum expectation of $\hat{V}_{ee}$. \cite{SeiGorSav-PRA-07} Physically, the solution of Eqs.~(\ref{eq_fradial})-(\ref{eq_fradialN}) makes the $N$ electrons always be in $N$ different circular shells, each of which contains, on average in the quantum mechanical problem (at $\is=1$), one electron. In the SCE limit, the electrons become strictly correlated, and all fluctuations are suppressed (see, e.g., \cite{ZieTaoSeiPer-IJQC-00}): the space is divided into $N$ regions, each of which always contains exactly one electron.
\subsection{The case $N=2$}
\label{sec_QDN2}
 In this case the minimizing angle is always $\theta_{12}(r)=\pi$ and there is only one co-motion function given by 
\beq
f_2(r)=\nf^{-1}(2-\nf(r)),
\eeq
with $f_2(f_2(r))=r$, thus ensuring the equivalence of the two electrons.

We switch to effective  Hartee units ($\hbar=1$, $a_B^*=\frac{\epsilon}{m^*} a_B=1$, $e=1$, $m^*=1$), and we define $f(r)\equiv f_2(r)$, so that
\beq
V_{\rm ee}^{\rm SCE}[\rho]=\int_0^{\infty} d r \,2\pi \,r\, \frac{\rho(r)}{N}\frac{1}{r+f(r)}=\int_0^{a_1} d r \,2\pi \,r\, \frac{\rho(r)}{r+f(r)},
\eeq
where we have used the fact that, since the electrons are indistinguishable, integrating from 0 to $\infty$ is equivalent to integrate $N$ times from 0 to $a_1=N_e^{-1}(1)$. This is a characteristic of the SCE limit: the space is divided in $N$ equivalent regions, so that to calculate the energy we only need to treat one of them. 
In a way, the SCE limit seems to become more ``local'', a characteristic which may prove very useful if we deal with approximations. However, we also have to keep in mind that, although for an exact evaluation of $V_{\rm ee}^{\rm SCE}[\rho]$ we need indeed only one of the $N$ equivalent regions, in order to find how to divide the space in those $N$ regions we need often to perform a classical minimization over the whole space. This will become clearer in the next example with $N=3$ electrons.

The exact ``first-order'' or zero-point energy is, in this case, given by
\beq
E_{\rm kd}^{\rm ZP}[\rho]=\int_0^{a_1}dr\,\pi r\,\rho(r) \left[\omega_1(r)+\omega_2(r)\right],
\eeq 
with
\begin{eqnarray}
\omega_1(r) & = & \sqrt{\frac{r^2+f(r)^2}{r f(r)\left(r+f(r)\right)^3}} \\
\omega_2(r) & = & \sqrt{\frac{2\left(1+f'(r)^2\right)}{-f'(r)\left(r+f(r)\right)^3}}	
\end{eqnarray}
\begin{figure}
  \begin{center}
   \includegraphics[width=13cm]{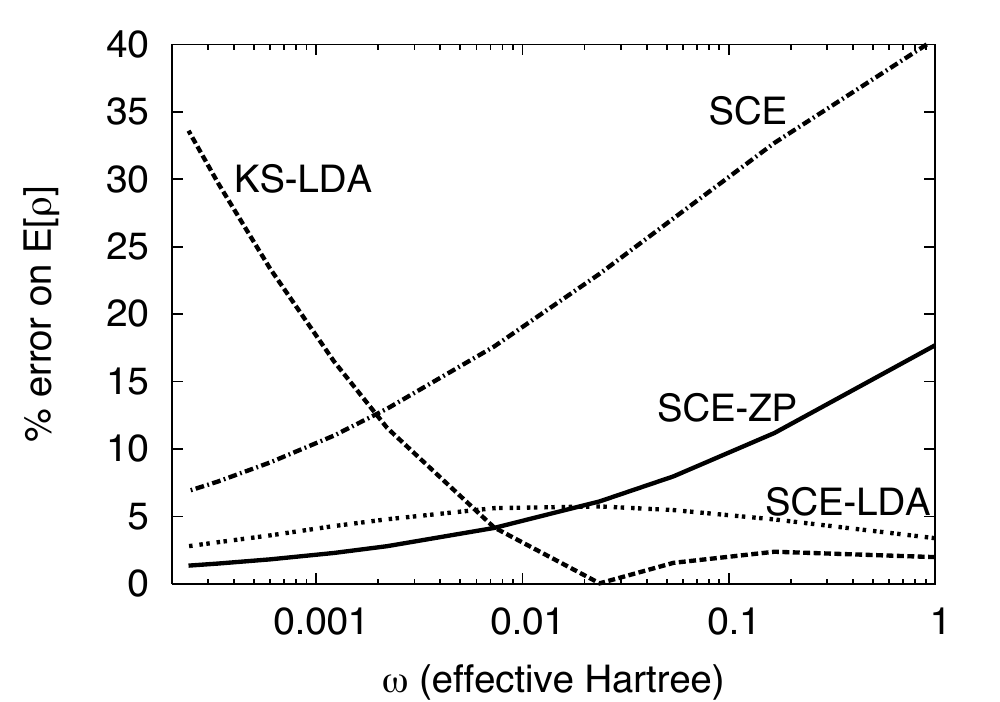}
   \caption{The absolute \% error on the total energy as a function of the confinement parameter $\omega$ made by the functional $E^{\rm ap}[v]=V_{\rm ee}^{\rm SCE}[\rho]+E_{\rm kd}^{\rm ap}[\rho]+\int d^Drv(\rv)\rho(\rv)$ with $E_{\rm kd}^{\rm ap}[\rho]=0$ (SCE), with $E_{\rm kd}^{\rm ap}[\rho]=E_{\rm kd}^{\rm LDA}[\rho]$ of Eq.~(\ref{eq_EkdLDA}) (SCE-LDA), and with $E_{\rm kd}^{\rm ap}[\rho]=E_{\rm kd}^{\rm ZP}[\rho]$ of Eq.~(\ref{eq_EkdZP}) (SCE-ZP). The results obtained with standard KS-LDA are also reported. In this figure all calculations are done at the postfunctional level only.}
\label{fig_errN2}
  \end{center}
\end{figure}

In Ref.~\onlinecite{GorSeiVig-PRL-09} we have evaluated the energy functional 
$E^{\rm ap}[v]=V_{\rm ee}^{\rm SCE}[\rho]+E_{\rm kd}^{\rm ap}[\rho]+
\int d^Drv(\rv)\rho(\rv)$ using the exact input densities from Ref.~\onlinecite{Tau-PAMG-94}, and we have compared the results with standard KS-LDA ones (notice that for two-dimensional electronic structure calculations LDA is still the most widely used functional). At this postfunctional level we have found that, as expected, for large values of the confining parameter $\omega$ (corresponding to higher densities) the KS LDA result is superior to the SCE-DFT. However, as $\omega$ becomes smaller (which corresponds to lowering the density and thus approaching the strongly-correlated regime), the SCE-DFT results with its approximations for $E_{\rm kd}^{\rm ap}[\rho]$ become better and better, highly outperforming KS-LDA. These results are summarized in Fig.~\ref{fig_errN2}, where we report the absolute \% error on the total energy as a function of the confinement parameter $\omega$ for KS-LDA and for SCE-DFT with $E_{\rm kd}^{\rm ap}[\rho]=0$  (curve labeled SCE), with $E_{\rm kd}^{\rm ap}[\rho]=E_{\rm kd}^{\rm LDA}[\rho]$ of Eq.~(\ref{eq_EkdLDA}) (SCE-LDA), and with $E_{\rm kd}^{\rm ap}[\rho]=E_{\rm kd}^{\rm ZP}[\rho]$ of Eq.~(\ref{eq_EkdZP}) (SCE-ZP). For the ground state energy of the 2D electron gas (which defines the LDA functional) we have used the data and parametrization of Attaccalite {\em et al.}\cite{AttMorGorBac-PRL-02} We see from Fig.~\ref{fig_errN2} that for $\omega\lesssim 0.007$ the SCE-ZP result is the most accurate. The much simpler SCE-LDA is also very reasonable in this regime, reducing the error of KS-LDA by a factor 5-10.

\begin{figure}
  \begin{center}
   \includegraphics[width=8.1cm]{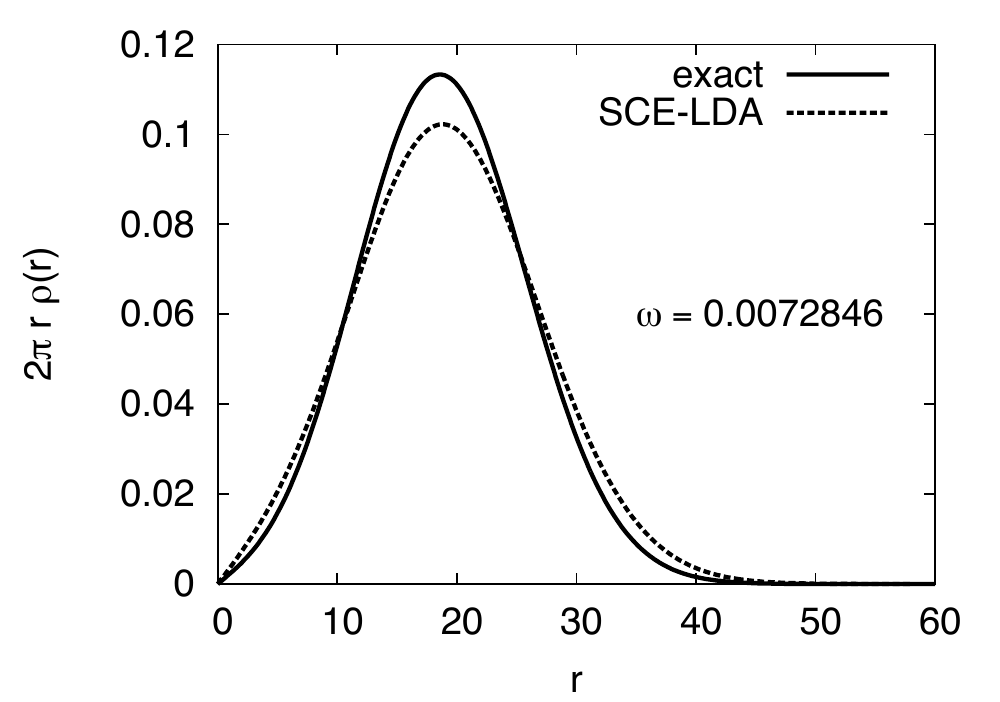}
\includegraphics[width=8.1cm]{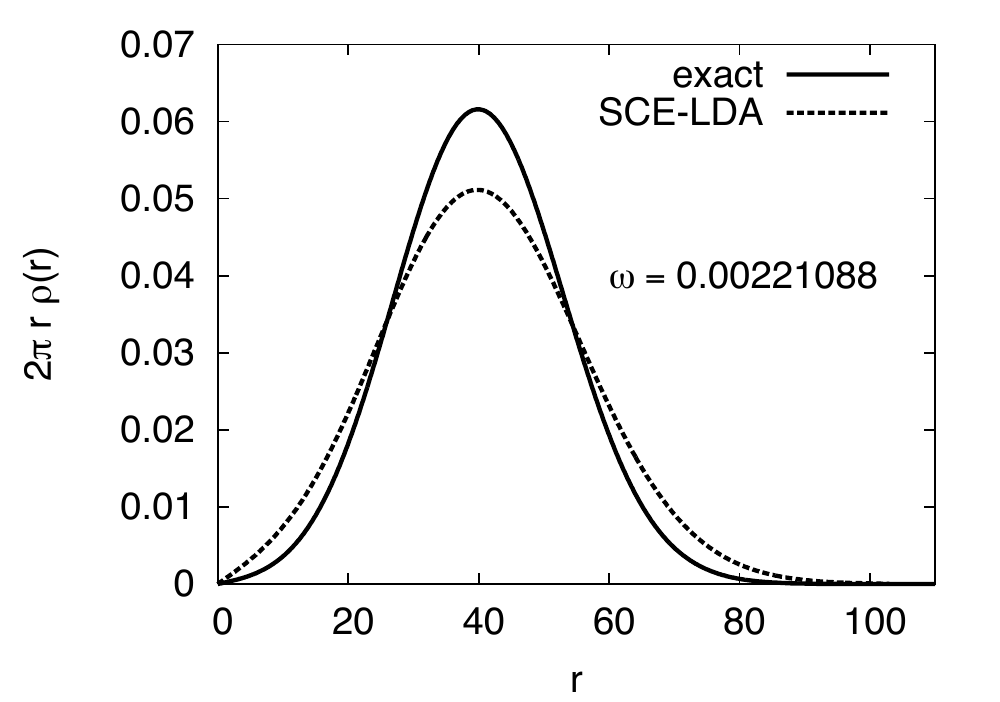}
   \caption{Radial densities for $N=2$ electrons in a two-dimensional model quantum dot for two different values of the confining parameter $\omega$. The exact values\cite{Tau-PAMG-94} are compared with the results obtained by the direct minimization of the energy functional SCE-LDA of Eq.~(\ref{eq_ESCEminPars}). Effective Hartree atomic units are used. The corresponding total energies have relative errors, respectively, of 5.4\%  and 4.4\%.}
\label{fig_densSC}
  \end{center}
\end{figure}

The next step is to perform self-consistent SCE-DFT calculations, in which the density is determined by minimizing the energy functional. Here we report very preliminary results obtained by parametrizing the density with a set of $N_g$ gaussians:
\beq
\rho_{\{p\}}(r)=C^{-1}\left(\sum_{i=1}^{N_g} c_i\, e^{-b_i^2\, r^2}\right)^2,
\eeq
where $\{p\}$ denotes the set of the $2N_g$ variational parameters $\{b_i,c_i;\;i=1,\dots N_g\}$. The constant $C$ ensures that $\rho(r)$ is normalized to $N=2$ electrons, and the functional form guarantees that $\rho(r)\ge 0$ everywhere. As an example, here we consider two cases with small confining parameter, $\omega=0.0072846$ and $\omega=0.00221088$, for which we find that $N_g=3$ gaussians are enough to accurately reproduce the exact density (when the fitted densities are inserted in $E^{\rm ap}[v]$ the error with respect to the energy obtained with the exact densities is $\sim 0.01 \%$). We consider only the simple SCE-LDA functional and perform the direct minimization
\beq
E^{\rm ap}[v]=\min_{\{p\}}\Big\{V_{\rm ee}^{\rm SCE}[\rho_{\{p\}}]+E_{\rm kd}^{\rm LDA}[\rho_{\{p\}}]+
\int d^Drv(\rv)\rho_{\{p\}}(\rv)\Big\}
\label{eq_ESCEminPars}
\eeq
with respect to the parameters $\{p\}$. This way of proceeding is probably not the best one both in terms of efficiency and accuracy, but the aim here is only to show a proof of principle. Better procedures are currently under study. The minimizing densities are compared in Fig.~\ref{fig_densSC} with the exact ones obtained from the solution given in Ref.~\onlinecite{Tau-PAMG-94}. Although the densities obtained are quite reasonable, it is evident that the LDA approximation for the functional $E_{\rm kd}[\rho]$ has a tendency to give densities that are too diffuse. The total energies obtained in this way are quite accurate, with errors of 5.4\% (for $\omega=0.0072846$) and 4.4\% (for $\omega=0.00221088$), corresponding, respectively, to absolute errors of 3~mH$^*$ and 1~mH$^*$.

\subsection{The case $N=3$}
\label{sec_QDN3}
In this case we have two co-motion functions, $f_2(r)$ and $f_3(r)$, and two relative angles that have to be minimized numerically for each value of the distance $r\in[0,a_1]$ of one of the electrons from the center of the dot. Notice that if, say, electron 1 is in the circular shell $0\le r\le a_1$, then electron 2 is in the shell $a_1\le f_2(r)\le a_2$, and electron 3 is in $a_2\le f_3(r)<\infty$. Thus, even if we only need to compute the minimizing angles for $r\in[0,a_1]$, we explore the whole space where $\rho(r)\neq 0$ through the positions of the other $N-1$ electrons.

The quantum dot with $N=3$ electrons is also a useful example to discuss the spin state in the framework of SCE-DFT. Accurate wavefunction methods, in fact, (see, e.g., \cite{RonCavBelGol-JCP-06,ZenGeiRuaUmrCho-PRB-09}) find that the ground state for the $N=3$ dot with $\omega\lesssim 0.05$ is fully spin polarized.
As discussed in Sec.~\ref{subsec_SCEapproach}, the functional $V_{ee}^{\rm SCE}[\rho]$, being essentially classic, is independent of the spin state. The exact functional $E_{\rm kd}[\rho]$ should thus be the same as the exact functional $E_{\rm kd}[\rho_\uparrow,\rho_\downarrow]$, when the exact density or the exact spin densities are used. When constructing approximations, however, one could obtain better results with $E_{\rm kd}[\rho_\uparrow,\rho_\downarrow]$, as in KS-DFT.

Here we consider only the SCE-LDA and SCE-LSD functionals, and we apply them at the postfunctional level using as input the Diffusion Monte Carlo densities from Refs.~\onlinecite{GhoGucUmrUllBar-NP-06,GucGhoUmrBar-PRB-08}. 
We study the values $\omega=0.01562$, 0.005 and 0.001, which already lie in the regime where KS-LDA orbitals become difficult to obtain (notice that the KS-LDA results of Fig.~\ref{fig_errN2} were obtained at the postfunctional level, using the exact densities as input).  As said, we explore the two options $E_{\rm kd}^{\rm LDA}[\rho]$ and $E_{\rm kd}^{\rm LSD}[\rho_\uparrow,\rho_\downarrow]$ for which we use the parametrization of the 2D electron gas energy of Attaccalite {\it et al.}\cite{AttMorGorBac-PRL-02} This functional is based on accurate Diffusion Monte Carlo (DMC) data predicting a weakly first order transition from the unpolarized gas to the fully polarized state at $r_s\approx 26$. Even if the existence of this transition has been recently questioned in Ref.~\onlinecite{DruNee-PRL-09}, we stick here to the original Attaccalite {\it et al.} parametrization. Since the densities involved are quite low, corresponding often to $r_s>26$, the correct definition (within the chosen parametrization) of the LDA functional consists in taking in each point of space the ground state energy of the electron gas with the same density, i.e.,
\beq
E_{\rm kd}^{\rm LDA}[\rho]=\int d^D r\,\rho(\rv)\left\{\epsilon_{\rm kd}\left(r_s(\rv),\zeta=0\right)\,\theta\left(25.56-r_s(\rv)\right)+\epsilon_{\rm kd}\left(r_s(\rv),\zeta=1\right)\,\theta\left(r_s(\rv)-25.56\right)\right\},
\label{eq_EkcLDAN3}
\eeq 
where $r_s(\rv)=(\pi \rho(\rv))^{-1/2}$, $\zeta=(\rho_\uparrow-\rho_\downarrow)/\rho$, and $\theta$ is the Heaviside step function. For the values of the confinement parameter $\omega$ considered here (for which the ground state of the dot is fully polarized), instead, the ``exact'' LSD functional (i.e., the one which has not only the exact local density in each point of space, but also the exact local spin densities) is
\beq
E_{\rm kd}^{\rm LSD}[\rho_\uparrow,0]=\int d^D r\,\rho(\rv)\epsilon_{\rm kd}(r_s(\rv),\zeta=1).
\label{eq_EkcLSDN3}
\eeq
\begin{table}
	\begin{center}
	\caption{Relative \% errors on the total energy of a model two-dimensional quantum dot consisting of 3 electrons confined in an harmonic potential $v_{\rm ext}(\rv)=\frac{1}{2}\omega^2 r^2$. Columns as follows: SCE are the results obtained by setting $E_{\rm kd}[\rho]=0$,  SCE-LDA are those obtained by using $E_{\rm kd}^{\rm LDA}[\rho]$ of Eq.~(\ref{eq_EkcLDAN3}), and SCE-LSD are those obtained by using $E_{\rm kd}^{\rm LSD}[\rho_\uparrow,0]$ of Eq.~(\ref{eq_EkcLSDN3}).}
	\label{tab_resN3}
\begin{tabular}{cccc}
\hline\hline
$\omega$                   & SCE   & SCE-LDA   & SCE-LSD   \\
\hline                                                    
0.01562    				  & $-15.1$  &  3.4      & 3.9   \\  
0.005     				 & $-10.6$  & 3.6       & 3.7   \\
0.001    				 & $-6.7$  & 2.8       & 2.8   \\

\hline\hline  
\end{tabular}
\end{center}
\end{table}

In Table~\ref{tab_resN3} we report the \% errors on the total energies (with respect to the DMC energies) obtained with the two functionals. We also show the results corresponding to $E_{\rm kd}^{\rm ap}[\rho]=0$, labeled ``SCE''. We see that the quality of the two local approximations is rather good, with the LSD results slightly worse than the LDA ones for $\omega=0.01562$ and $\omega=0.005$. This is due to the fact that for these values of the confining parameter $\omega$, $r_s(\rv)$ is often still smaller that $26$, so that a lower energy is obtained by considering the true ground state of the electron gas. At $\omega=0.001$, we have $r_s(\rv)$ always greater than 26 so that LDA and LSD become the same. In other words, the SCE-LDA functional predicts a transition to the fully polarized state at a much lower $\omega$ with respect to the one predicted by accurate wavefunction methods. This transition in the SCE-LDA method entirely depends on the delicate physics of the 2D uniform electron gas, and it is thus questionable in view of the latest results of Ref.~\onlinecite{DruNee-PRL-09}.

This simple example shows that the next step for the construction of  functionals useful for SCE-DFT is probably by considering simple exchange models, which would allow to distinguish between different spin states, generalizing to nonuniform densities what has been done for the uniform electron gas in Ref.~\onlinecite{Car-PR-61}.

\section{Is the SCE limit relevant for chemical applications?}
\label{sec_SCEchem}
The results of the previous Section suggest that the SCE formalism can have an impact on solid-state devices 
involving electron gas in low dimensional systems (quantum wires, dots, point contacts, 
etc.), in the low-density, strongly-interacting regime, where traditional KS DFT is not of much use.
It is however less evident whether the SCE limit could be also relevant for applications in chemistry.

If we consider the simplest chemical system, the H$_2$ molecule, we see that, as we stretch the chemical bond, the energy and physics of the system is exactly described by the SCE limit, as electrons in a stretched bond have strong spatial correlations (see, e.g., Fig.~11 of Ref.~\onlinecite{TeaCorHel-JCP-10}).
This feature is very interesting and promising, since the stretching of the chemical bond is one of the typical situations in which restricted KS-DFT encounters problems, being unable to describe the strong correlation occurring between the electrons involved in a single or in a multiple bond. Thus, the SCE limit contains useful exact information that is usually missed by state-of-the-art (restricted) KS-DFT. 
However, when we deal with real chemical systems the situation is different from that of the simple H$_2$ molecule, since only the electrons involved in the stretched bonds are strongly correlated. The SCE limit applied to the whole system would give much too low energies, producing serious overcorrelation. In other words, we cannot expect the SCE-DFT scheme to work for chemistry, where often both the orbital description and strong spatial correlation are important at the same time. 

What we could do, instead, is trying to include the exact information contained in the SCE limit into approximate exchange-correlation functionals. Attempts in this direction have been done in the past, leading to the construction of the interaction-strength-interpolation (ISI) functional.\cite{SeiPerLev-PRA-99,SeiPerKur-PRL-00} As shown in Eq.~(\ref{eq_ExcH}), the exchange-correlation energy of KS-DFT is given by (in this section we use Hartree atomic units)
\beq
E_{\rm xc}[\rho]=\int_0^1d\is V_{\rm ee}^{(\is)}[\rho]-U[\rho].
\label{eq_adiasolita}
\eeq
Since  the functional $V_{\rm ee}^{(\is)}[\rho]$ approaches the SCE limit as $\is\to\infty$, the idea of the ISI functional is to construct the $\is$-dependence of $W_\is[\rho]=V_{\rm ee}^{(\is)}[\rho]-U[\rho]$ by interpolating between the $\is\to 0$ (exchange energy and second-order G\"orling-Levy perturbation energy\cite{GorLev-PRA-94} $E_c^{\rm GL2}[\rho]$),
\beq
W_{\is\to 0}[\rho]=V_{\rm ee}^{(\is\to 0)}[\rho]-U[\rho]=E_x[\rho]+2\,\is \,E_c^{\rm GL2}[\rho]+O(\is^2),
\label{eq_Wisto0}
\eeq
 and the $\is\to\infty$ limits (SCE plus ZP oscillations\cite{GorVigSei-JCTC-09}),
\beq
W_{\is\to \infty}[\rho]=V_{\rm ee}^{(\is\to \infty)}[\rho]-U[\rho]=V_{\rm ee}^{\rm SCE}[\rho]-U[\rho]+\frac{T_{\rm ZP}[\rho] }{\sqrt{\is}}+O(\is^{-q}) \qquad q\ge\frac{5}{4}.
\label{eq_Wistoinfty}
\eeq
However, this way of proceeding leads to serious size-consistency errors. The size-consistency problem of the ISI functional is related to the fact that the interpolation is done on the global quantity $W_\is[\rho]$. Moreover, when the ISI was first proposed an exact treatment of the SCE limit was not available, so that the functional relied on physical approximations for the SCE and ZP energies.\cite{SeiPerKur-PRA-00,SeiPerKur-PRL-00} 

As a possible way out, the exact solution of the SCE limit, now available, makes accessible not only global, but also {\em local} quantities. This new access to local quantities could be used to construct {\it local} interpolations along the DFT adiabatic connection, restoring size consistency (for critical reviews on the size-consistency issue in DFT see also\cite{GorSav-JPCS-08,Sav-CP-09}). We thus rewrite Eq.~(\ref{eq_adiasolita}) in terms of an energy density $w_{\is}(\rv;[\rho])$,
\beq
E_{\rm xc}[\rho]=\int d^D r \rho(\rv)\int_0^1d\is\, w_{\is}(\rv;[\rho]),
\eeq
with
\beq
\int d^D r \rho(\rv) w_{\is}(\rv;[\rho])=W_\is[\rho]= V_{\rm ee}^{(\is)}[\rho]-U[\rho].
\eeq
The idea is then to use the energy densities $w_{\is}(\rv;[\rho])$ in the $\is\to 0$ and $\is\to\infty$ limits, describing locally the quantities of Eqs.~(\ref{eq_Wisto0})-(\ref{eq_Wistoinfty}), in order to construct an interpolation for the $\is-$dependence of $w_{\is}(\rv;[\rho])$.
Since the energy density $w_{\is}(\rv;[\rho])$ is not uniquely defined, we must use the same gauge for the weak and and the strong-interaction limits.
 A very reasonable and physical choice would be the gauge defined by the exchange-correlation hole,
\beq
w_{\is}(\rv,[\rho])=\frac{1}{2}\int d^D u \frac{\rho_{\rm xc}^\is(\rv,u)}{u},
\eeq
where $\uv=\rv_2-\rv_1$, $u=|\uv|$ and the exchange-correlation hole $\rho_{\rm xc}^\is(\rv,u)$ is simply related to the pair density $P_2^\is(\rv_1,\rv_2)$ obtained from the wavefunction $\Psi_\is$,
\begin{eqnarray}
P_2^\is(\rv_1,\rv_2) & = & N(N-1)\sum_{\sigma_1,\dots,\sigma_2}\int d^D r_3\dots d^D r_N |\Psi_\is(\rv_1,\sigma_1,\dots\rv_N,\sigma_N)|^2, \\
\rho_{\rm xc}^\is(\rv,u) & = & \frac{1}{\rho(\rv)}\int \frac{d\hat{\uv}}{4\pi}\left(P_2^\is(\rv,\rv+\uv)-\rho(\rv)\rho(\rv+\uv)\right).
\end{eqnarray}
The $\is\to 0$ limit of $w_{\is}(\rv;[\rho])$ is thus the exchange energy density defined in the gauge of the exchange hole, for which one could use the exact exchange hole or a good approximation, e.g., the one of Becke and Roussel.\cite{BecRou-PRA-89} The $\is\to \infty$ limit of $w_{\is}(\rv;[\rho])$ is exactly given by the SCE solution, which is already defined in the gauge of the exchange-correlation hole (see also Ref.~\onlinecite{GorSeiSav-PCCP-08}),
\beq
w_{\is\to\infty}(\rv,[\rho])=\frac{1}{N}\sum_{i,j=1}^N\frac{1-\delta_{ij}}{|\fv_i(\rv)-\fv_j(\rv)|}-\int \frac{d^D u}{u}\rho(\rv+\uv).
\eeq
Much more difficult is to have a local expression for the next leading terms, both for $\is\to 0$ and $\is\to\infty$, defined in the same gauge. The zero-point term of Eq.~(\ref{eq_EkdZP}), which determines how the $\is\to\infty$ limit is approached to orders $\is^{-1/2}$, is, in fact, expressed in a gauge which is not the one of the exchange-correlation (xc) hole. The G\"orling-Levy perturbation theory is also difficult to define locally in terms of the xc-hole gauge. 

Routes to define and calculate the local next leading terms will be pursued in future work. For the ZP term, one could actually directly calculate the pair-density associated to the $O(\is^{-1/2})$ wavefunction,\cite{GorVigSei-JCTC-09} and produce the exact exchange-correlation hole in this limit. For the $\is\to 0$ leading correction, one should probably use different correlation-strength indicators than the GL perturbation theory. 
A very promising route could be the one described by Becke in Ref.~\onlinecite{Bec-JCP-03}, which considers the local normalization of the exact exchange hole as an indicator of strong non-dynamical correlation.

The main message of this Section is that the SCE limit contains useful exact information for critical situations in Chemistry such as stretched bonds. However, one has to be able to use this exact information locally, where it is needed. This direction of research will be pursued in future work.

\section{Concluding remarks}
\label{sec_conc}
The strong-interaction limit of density functional theory, exactly solved in the last three years, contains useful physical and chemical information, typically missed by standard Kohn-Sham DFT. In this paper we have outlined some paths to fully exploit this piece of exact information, with the aim of broadening the applicability of DFT for electronic structure calculations in solid-state physical devices and in chemical systems, addressing fundamental issues of standard KS DFT.

The mathematical structure of the strong-interaction limit of DFT has been uncovered in Refs.~\onlinecite{SeiGorSav-PRA-07,GorSeiSav-PCCP-08,GorVigSei-JCTC-09}. However, solving the relevant equations for a general density in an efficient way is still an open problem, which will be addressed in future work, exploiting the formal similarity with mass transportation theory.\cite{ButDepGor-PRA-10}

Another line of research for future work is based on the fact that the strictly correlated problem defined by the strong-interaction limit of DFT provides a physical, rigorous, lower bound for the exact exchange-correlation functional of standard Kohn-Sham DFT, a feature which may be exploited for the construction of approximate functionals.\cite{RasSeiGor-XXX-10} 

The calculation and study of energy densities in the strong-interaction limit of DFT will also provide useful information to be included into approximate functionals, and will be the object of future work.
 
\section*{Acknowledgments}
We thank Cyrus Umrigar and Devrim Guclu for the densities of the $N=3$ quantum dots. This work was supported by the Netherlands Organization for Scientific Research (NWO) through a Vidi grant.


\begin{thebibliography}{81}
\expandafter\ifx\csname natexlab\endcsname\relax\def\natexlab#1{#1}\fi
\expandafter\ifx\csname bibnamefont\endcsname\relax
  \def\bibnamefont#1{#1}\fi
\expandafter\ifx\csname bibfnamefont\endcsname\relax
  \def\bibfnamefont#1{#1}\fi
\expandafter\ifx\csname citenamefont\endcsname\relax
  \def\citenamefont#1{#1}\fi
\expandafter\ifx\csname url\endcsname\relax
  \def\url#1{\texttt{#1}}\fi
\expandafter\ifx\csname urlprefix\endcsname\relax\def\urlprefix{URL }\fi
\providecommand{\bibinfo}[2]{#2}
\providecommand{\eprint}[2][]{\url{#2}}

\bibitem[{\citenamefont{Kohn}(1999)}]{Koh-RMP-99}
\bibinfo{author}{\bibfnamefont{W.}~\bibnamefont{Kohn}}, \bibinfo{journal}{Rev.
  Mod. Phys.} \textbf{\bibinfo{volume}{{71}}}, \bibinfo{pages}{1253}
  (\bibinfo{year}{1999}).

\bibitem[{\citenamefont{Kohn and Sham}(1965)}]{KohSha-PR-65}
\bibinfo{author}{\bibfnamefont{W.}~\bibnamefont{Kohn}} \bibnamefont{and}
  \bibinfo{author}{\bibfnamefont{L.~J.} \bibnamefont{Sham}},
  \bibinfo{journal}{Phys. Rev. A} \textbf{\bibinfo{volume}{140}},
  \bibinfo{pages}{1133} (\bibinfo{year}{1965}).

\bibitem[{\citenamefont{Runge and Gross}(1984)}]{RunGro-PRL-84}
\bibinfo{author}{\bibfnamefont{E.}~\bibnamefont{Runge}} \bibnamefont{and}
  \bibinfo{author}{\bibfnamefont{E.~K.~U.} \bibnamefont{Gross}},
  \bibinfo{journal}{Phys. Rev. Lett.} \textbf{\bibinfo{volume}{52}},
  \bibinfo{pages}{997} (\bibinfo{year}{1984}).

\bibitem[{\citenamefont{Mattsson}(2002)}]{Mat-SCI-02}
\bibinfo{author}{\bibfnamefont{A.~E.} \bibnamefont{Mattsson}},
  \bibinfo{journal}{Science} \textbf{\bibinfo{volume}{{298}}},
  \bibinfo{pages}{759} (\bibinfo{year}{2002}).

\bibitem[{\citenamefont{Perdew et~al.}(2005)\citenamefont{Perdew, Ruzsinszky,
  Tao, Staroverov, Scuseria, and Csonka}}]{PerRuzTaoStaScuCso-JCP-05}
\bibinfo{author}{\bibfnamefont{J.~P.} \bibnamefont{Perdew}},
  \bibinfo{author}{\bibfnamefont{A.}~\bibnamefont{Ruzsinszky}},
  \bibinfo{author}{\bibfnamefont{J.}~\bibnamefont{Tao}},
  \bibinfo{author}{\bibfnamefont{V.~N.} \bibnamefont{Staroverov}},
  \bibinfo{author}{\bibfnamefont{G.~E.} \bibnamefont{Scuseria}},
  \bibnamefont{and} \bibinfo{author}{\bibfnamefont{G.~I.}
  \bibnamefont{Csonka}}, \bibinfo{journal}{J. Chem. Phys.}
  \textbf{\bibinfo{volume}{123}}, \bibinfo{pages}{062201}
  (\bibinfo{year}{2005}).

\bibitem[{\citenamefont{Becke and Johnson}(2007)}]{BecJoh-JCP-07}
\bibinfo{author}{\bibfnamefont{A.~D.} \bibnamefont{Becke}} \bibnamefont{and}
  \bibinfo{author}{\bibfnamefont{E.~R.} \bibnamefont{Johnson}},
  \bibinfo{journal}{J. Chem. Phys.} \textbf{\bibinfo{volume}{{127}}},
  \bibinfo{pages}{124108} (\bibinfo{year}{2007}).

\bibitem[{\citenamefont{Zhao et~al.}(2006)\citenamefont{Zhao, Schultz, and
  Truhlar}}]{ZhaSchTru-JCTC-06}
\bibinfo{author}{\bibfnamefont{Y.}~\bibnamefont{Zhao}},
  \bibinfo{author}{\bibfnamefont{N.~E.} \bibnamefont{Schultz}},
  \bibnamefont{and} \bibinfo{author}{\bibfnamefont{D.~G.}
  \bibnamefont{Truhlar}}, \bibinfo{journal}{J. Chem. Theory Comput.}
  \textbf{\bibinfo{volume}{2}}, \bibinfo{pages}{364} (\bibinfo{year}{2006}).

\bibitem[{\citenamefont{Cohen et~al.}(2008)\citenamefont{Cohen, Mori-Sanchez,
  and Yang}}]{CohMorYan-SCI-08}
\bibinfo{author}{\bibfnamefont{A.~J.} \bibnamefont{Cohen}},
  \bibinfo{author}{\bibfnamefont{P.}~\bibnamefont{Mori-Sanchez}},
  \bibnamefont{and} \bibinfo{author}{\bibfnamefont{W.~T.} \bibnamefont{Yang}},
  \bibinfo{journal}{Science} \textbf{\bibinfo{volume}{{321}}},
  \bibinfo{pages}{792} (\bibinfo{year}{2008}).

\bibitem[{\citenamefont{Savin}(1996)}]{Sav-INC-96}
\bibinfo{author}{\bibfnamefont{A.}~\bibnamefont{Savin}}, in
  \emph{\bibinfo{booktitle}{Recent Developments of Modern Density Functional
  Theory}}, edited by \bibinfo{editor}{\bibfnamefont{J.~M.}
  \bibnamefont{Seminario}} (\bibinfo{publisher}{Elsevier},
  \bibinfo{address}{Amsterdam}, \bibinfo{year}{1996}), pp.
  \bibinfo{pages}{327--357}.

\bibitem[{\citenamefont{Leininger et~al.}(1997)\citenamefont{Leininger, Stoll,
  Werner, and Savin}}]{LeiStoWerSav-CPL-97}
\bibinfo{author}{\bibfnamefont{T.}~\bibnamefont{Leininger}},
  \bibinfo{author}{\bibfnamefont{H.}~\bibnamefont{Stoll}},
  \bibinfo{author}{\bibfnamefont{H.-J.} \bibnamefont{Werner}},
  \bibnamefont{and} \bibinfo{author}{\bibfnamefont{A.}~\bibnamefont{Savin}},
  \bibinfo{journal}{Chem. Phys. Lett.} \textbf{\bibinfo{volume}{{275}}},
  \bibinfo{pages}{151} (\bibinfo{year}{1997}).

\bibitem[{\citenamefont{Pollet et~al.}(2002)\citenamefont{Pollet, Savin,
  Leininger, and Stoll}}]{PolSavLeiSto-JCP-02}
\bibinfo{author}{\bibfnamefont{R.}~\bibnamefont{Pollet}},
  \bibinfo{author}{\bibfnamefont{A.}~\bibnamefont{Savin}},
  \bibinfo{author}{\bibfnamefont{T.}~\bibnamefont{Leininger}},
  \bibnamefont{and} \bibinfo{author}{\bibfnamefont{H.}~\bibnamefont{Stoll}},
  \bibinfo{journal}{J. Chem. Phys.} \textbf{\bibinfo{volume}{{116}}},
  \bibinfo{pages}{1250} (\bibinfo{year}{2002}).

\bibitem[{\citenamefont{\'Angy\'an et~al.}(2005)\citenamefont{\'Angy\'an,
  Gerber, Savin, and Toulouse}}]{AngGerSavTou-PRA-05}
\bibinfo{author}{\bibfnamefont{J.~G.} \bibnamefont{\'Angy\'an}},
  \bibinfo{author}{\bibfnamefont{I.}~\bibnamefont{Gerber}},
  \bibinfo{author}{\bibfnamefont{A.}~\bibnamefont{Savin}}, \bibnamefont{and}
  \bibinfo{author}{\bibfnamefont{J.}~\bibnamefont{Toulouse}},
  \bibinfo{journal}{Phys. Rev. A} \textbf{\bibinfo{volume}{{72}}},
  \bibinfo{pages}{012510} (\bibinfo{year}{2005}).

\bibitem[{\citenamefont{Goll et~al.}(2005)\citenamefont{Goll, Werner, and
  Stoll}}]{GolWerSto-PCCP-05}
\bibinfo{author}{\bibfnamefont{E.}~\bibnamefont{Goll}},
  \bibinfo{author}{\bibfnamefont{H.-J.} \bibnamefont{Werner}},
  \bibnamefont{and} \bibinfo{author}{\bibfnamefont{H.}~\bibnamefont{Stoll}},
  \bibinfo{journal}{Phys. Chem. Chem. Phys.} \textbf{\bibinfo{volume}{7}},
  \bibinfo{pages}{3917} (\bibinfo{year}{2005}).

\bibitem[{\citenamefont{Goll et~al.}(2006)\citenamefont{Goll, Werner, Stoll,
  Leininger, Gori-Giorgi, and Savin}}]{GolWerStoLeiGorSav-CP-06}
\bibinfo{author}{\bibfnamefont{E.}~\bibnamefont{Goll}},
  \bibinfo{author}{\bibfnamefont{H.-J.} \bibnamefont{Werner}},
  \bibinfo{author}{\bibfnamefont{H.}~\bibnamefont{Stoll}},
  \bibinfo{author}{\bibfnamefont{T.}~\bibnamefont{Leininger}},
  \bibinfo{author}{\bibfnamefont{P.}~\bibnamefont{Gori-Giorgi}},
  \bibnamefont{and} \bibinfo{author}{\bibfnamefont{A.}~\bibnamefont{Savin}},
  \bibinfo{journal}{Chem. Phys.} \textbf{\bibinfo{volume}{329}},
  \bibinfo{pages}{276} (\bibinfo{year}{2006}).

\bibitem[{\citenamefont{Fromager et~al.}(2007)\citenamefont{Fromager, Toulouse,
  and Jensen}}]{FroTouJen-JCP-07}
\bibinfo{author}{\bibfnamefont{E.}~\bibnamefont{Fromager}},
  \bibinfo{author}{\bibfnamefont{J.}~\bibnamefont{Toulouse}}, \bibnamefont{and}
  \bibinfo{author}{\bibfnamefont{H.~J.~A.} \bibnamefont{Jensen}},
  \bibinfo{journal}{J. Chem. Phys.} \textbf{\bibinfo{volume}{126}},
  \bibinfo{pages}{074111} (\bibinfo{year}{2007}).

\bibitem[{\citenamefont{Toulouse et~al.}(2009)\citenamefont{Toulouse, Gerber,
  Jansen, Savin, and \'Angy\'an}}]{TouGerJanSavAng-PRL-09}
\bibinfo{author}{\bibfnamefont{J.}~\bibnamefont{Toulouse}},
  \bibinfo{author}{\bibfnamefont{I.~C.} \bibnamefont{Gerber}},
  \bibinfo{author}{\bibfnamefont{G.}~\bibnamefont{Jansen}},
  \bibinfo{author}{\bibfnamefont{A.}~\bibnamefont{Savin}}, \bibnamefont{and}
  \bibinfo{author}{\bibfnamefont{J.~G.} \bibnamefont{\'Angy\'an}},
  \bibinfo{journal}{Phys. Rev. Lett.} \textbf{\bibinfo{volume}{{102}}},
  \bibinfo{pages}{096404} (\bibinfo{year}{2009}).

\bibitem[{\citenamefont{Janesko et~al.}(2009)\citenamefont{Janesko, Henderson,
  and Scuseria}}]{JanHenScu-JCP-09}
\bibinfo{author}{\bibfnamefont{B.~G.} \bibnamefont{Janesko}},
  \bibinfo{author}{\bibfnamefont{T.~M.} \bibnamefont{Henderson}},
  \bibnamefont{and} \bibinfo{author}{\bibfnamefont{G.~E.}
  \bibnamefont{Scuseria}}, \bibinfo{journal}{J. Chem. Phys.}
  \textbf{\bibinfo{volume}{{130}}}, \bibinfo{pages}{081105}
  (\bibinfo{year}{2009}).

\bibitem[{\citenamefont{Livshits and Baer}(2007)}]{LivBae-PCCP-07}
\bibinfo{author}{\bibfnamefont{E.}~\bibnamefont{Livshits}} \bibnamefont{and}
  \bibinfo{author}{\bibfnamefont{R.}~\bibnamefont{Baer}},
  \bibinfo{journal}{Phys. Chem. Chem. Phys.} \textbf{\bibinfo{volume}{9}},
  \bibinfo{pages}{2932} (\bibinfo{year}{2007}).

\bibitem[{\citenamefont{Goll et~al.}(2007)\citenamefont{Goll, Stoll,
  Thierfelder, and Schwerdtfeger}}]{GolStoThiSch-PRA-07}
\bibinfo{author}{\bibfnamefont{E.}~\bibnamefont{Goll}},
  \bibinfo{author}{\bibfnamefont{H.}~\bibnamefont{Stoll}},
  \bibinfo{author}{\bibfnamefont{C.}~\bibnamefont{Thierfelder}},
  \bibnamefont{and}
  \bibinfo{author}{\bibfnamefont{P.}~\bibnamefont{Schwerdtfeger}},
  \bibinfo{journal}{Phys. Rev. A} \textbf{\bibinfo{volume}{76}},
  \bibinfo{pages}{032507} (\bibinfo{year}{2007}).

\bibitem[{\citenamefont{Goll et~al.}(2008)\citenamefont{Goll, Leininger, Manby,
  Mitrushchenkov, Werner, and Stoll}}]{GolLeiManMitWerSto-PCCP-08}
\bibinfo{author}{\bibfnamefont{E.}~\bibnamefont{Goll}},
  \bibinfo{author}{\bibfnamefont{T.}~\bibnamefont{Leininger}},
  \bibinfo{author}{\bibfnamefont{F.~R.} \bibnamefont{Manby}},
  \bibinfo{author}{\bibfnamefont{A.}~\bibnamefont{Mitrushchenkov}},
  \bibinfo{author}{\bibfnamefont{H.-J.} \bibnamefont{Werner}},
  \bibnamefont{and} \bibinfo{author}{\bibfnamefont{H.}~\bibnamefont{Stoll}},
  \bibinfo{journal}{Phys. Chem. Chem. Phys.} \textbf{\bibinfo{volume}{10}},
  \bibinfo{pages}{3353} (\bibinfo{year}{2008}).

\bibitem[{\citenamefont{Fromager et~al.}(2010)\citenamefont{Fromager,
  Cimiraglia, and Jensen}}]{FroCimJen-PRA-10}
\bibinfo{author}{\bibfnamefont{E.}~\bibnamefont{Fromager}},
  \bibinfo{author}{\bibfnamefont{R.}~\bibnamefont{Cimiraglia}},
  \bibnamefont{and} \bibinfo{author}{\bibfnamefont{H.~J.~A.}
  \bibnamefont{Jensen}}, \bibinfo{journal}{Phys. Rev. A}
  \textbf{\bibinfo{volume}{{81}}}, \bibinfo{pages}{024502}
  (\bibinfo{year}{2010}).

\bibitem[{\citenamefont{Paier et~al.}(2010)\citenamefont{Paier, Janesko,
  Henderson, Scuseria, Gr\"uneis, and Kresse}}]{PaiJanHenScuGruKre-JCP-10}
\bibinfo{author}{\bibfnamefont{J.}~\bibnamefont{Paier}},
  \bibinfo{author}{\bibfnamefont{B.~G.} \bibnamefont{Janesko}},
  \bibinfo{author}{\bibfnamefont{T.~M.} \bibnamefont{Henderson}},
  \bibinfo{author}{\bibfnamefont{G.~E.} \bibnamefont{Scuseria}},
  \bibinfo{author}{\bibfnamefont{A.}~\bibnamefont{Gr\"uneis}},
  \bibnamefont{and} \bibinfo{author}{\bibfnamefont{G.}~\bibnamefont{Kresse}},
  \bibinfo{journal}{J. Chem. Phys.} \textbf{\bibinfo{volume}{{132}}},
  \bibinfo{pages}{094103} (\bibinfo{year}{2010}).

\bibitem[{\citenamefont{Zhu et~al.}(2010)\citenamefont{Zhu, Toulouse, Savin,
  and \'Angy\'an}}]{ZhuTouSavAng-JCP-10}
\bibinfo{author}{\bibfnamefont{W.}~\bibnamefont{Zhu}},
  \bibinfo{author}{\bibfnamefont{J.}~\bibnamefont{Toulouse}},
  \bibinfo{author}{\bibfnamefont{A.}~\bibnamefont{Savin}}, \bibnamefont{and}
  \bibinfo{author}{\bibfnamefont{J.~G.} \bibnamefont{\'Angy\'an}},
  \bibinfo{journal}{J. Chem. Phys.} \textbf{\bibinfo{volume}{{132}}},
  \bibinfo{pages}{244108} (\bibinfo{year}{2010}).

\bibitem[{\citenamefont{Cioslowski and Pernal}(2000)}]{CioPer-JCP-00}
\bibinfo{author}{\bibfnamefont{J.}~\bibnamefont{Cioslowski}} \bibnamefont{and}
  \bibinfo{author}{\bibfnamefont{K.}~\bibnamefont{Pernal}},
  \bibinfo{journal}{J. Chem. Phys.} \textbf{\bibinfo{volume}{{113}}},
  \bibinfo{pages}{8434} (\bibinfo{year}{2000}).

\bibitem[{\citenamefont{Cioslowski and Buchowiecki}(2006)}]{CioBuc-JCP-06}
\bibinfo{author}{\bibfnamefont{J.}~\bibnamefont{Cioslowski}} \bibnamefont{and}
  \bibinfo{author}{\bibfnamefont{M.}~\bibnamefont{Buchowiecki}},
  \bibinfo{journal}{J. Chem. Phys.} \textbf{\bibinfo{volume}{{125}}},
  \bibinfo{pages}{064105} (\bibinfo{year}{2006}).

\bibitem[{\citenamefont{Zhitenev et~al.}(1997)\citenamefont{Zhitenev, Ashoori,
  Pfeiffer, and West}}]{ZhiAshPfeWes-PRL-97}
\bibinfo{author}{\bibfnamefont{N.~B.} \bibnamefont{Zhitenev}},
  \bibinfo{author}{\bibfnamefont{R.~C.} \bibnamefont{Ashoori}},
  \bibinfo{author}{\bibfnamefont{L.~N.} \bibnamefont{Pfeiffer}},
  \bibnamefont{and} \bibinfo{author}{\bibfnamefont{K.~W.} \bibnamefont{West}},
  \bibinfo{journal}{Phys. Rev. Lett.} \textbf{\bibinfo{volume}{79}},
  \bibinfo{pages}{2308} (\bibinfo{year}{1997}).

\bibitem[{\citenamefont{Gritsenko et~al.}(2005)\citenamefont{Gritsenko, Pernal,
  and Baerends}}]{GriPerBae-JCP-05}
\bibinfo{author}{\bibfnamefont{O.}~\bibnamefont{Gritsenko}},
  \bibinfo{author}{\bibfnamefont{K.}~\bibnamefont{Pernal}}, \bibnamefont{and}
  \bibinfo{author}{\bibfnamefont{E.~J.} \bibnamefont{Baerends}},
  \bibinfo{journal}{J. Chem. Phys.} \textbf{\bibinfo{volume}{{122}}},
  \bibinfo{pages}{204102} (\bibinfo{year}{2005}).

\bibitem[{\citenamefont{Rohr et~al.}(2008)\citenamefont{Rohr, Pernal,
  Gritsenko, and Baerends}}]{RohPerGriBae-JCP-08}
\bibinfo{author}{\bibfnamefont{D.~R.} \bibnamefont{Rohr}},
  \bibinfo{author}{\bibfnamefont{K.}~\bibnamefont{Pernal}},
  \bibinfo{author}{\bibfnamefont{O.~V.} \bibnamefont{Gritsenko}},
  \bibnamefont{and} \bibinfo{author}{\bibfnamefont{E.~J.}
  \bibnamefont{Baerends}}, \bibinfo{journal}{J. Chem. Phys.}
  \textbf{\bibinfo{volume}{{129}}}, \bibinfo{pages}{164105}
  (\bibinfo{year}{2008}).

\bibitem[{\citenamefont{Tsuchimochi and Scuseria}(2009)}]{TsuScu-JCP-09}
\bibinfo{author}{\bibfnamefont{T.}~\bibnamefont{Tsuchimochi}} \bibnamefont{and}
  \bibinfo{author}{\bibfnamefont{G.~E.} \bibnamefont{Scuseria}}
  (\bibinfo{year}{2009}).

\bibitem[{\citenamefont{Levy}(1979)}]{Lev-PNAS-79}
\bibinfo{author}{\bibfnamefont{M.}~\bibnamefont{Levy}}, \bibinfo{journal}{Proc.
  Natl. Acad. Sci. U.S.A.} \textbf{\bibinfo{volume}{76}}, \bibinfo{pages}{6062}
  (\bibinfo{year}{1979}).

\bibitem[{\citenamefont{Levy and Perdew}(1985{\natexlab{a}})}]{LevPer-INC-85}
\bibinfo{author}{\bibfnamefont{M.}~\bibnamefont{Levy}} \bibnamefont{and}
  \bibinfo{author}{\bibfnamefont{J.~P.} \bibnamefont{Perdew}}, in
  \emph{\bibinfo{booktitle}{Density Functional Methods in Physics}}, edited by
  \bibinfo{editor}{\bibfnamefont{R.~M.} \bibnamefont{Dreizler}}
  \bibnamefont{and}
  \bibinfo{editor}{\bibfnamefont{J.}~\bibnamefont{da~Providencia}}
  (\bibinfo{publisher}{Plenum}, \bibinfo{address}{New York},
  \bibinfo{year}{1985}{\natexlab{a}}).

\bibitem[{\citenamefont{Harris}(1984)}]{Har-PRA-84}
\bibinfo{author}{\bibfnamefont{J.}~\bibnamefont{Harris}},
  \bibinfo{journal}{Phys. Rev. A} \textbf{\bibinfo{volume}{{29}}},
  \bibinfo{pages}{1648} (\bibinfo{year}{1984}).

\bibitem[{\citenamefont{Langreth and Perdew}(1975)}]{LanPer-SSC-75}
\bibinfo{author}{\bibfnamefont{D.~C.} \bibnamefont{Langreth}} \bibnamefont{and}
  \bibinfo{author}{\bibfnamefont{J.~P.} \bibnamefont{Perdew}},
  \bibinfo{journal}{Solid State Commun.} \textbf{\bibinfo{volume}{{17}}},
  \bibinfo{pages}{1425} (\bibinfo{year}{1975}).

\bibitem[{\citenamefont{Yang}(1998)}]{Yan-JCP-98}
\bibinfo{author}{\bibfnamefont{W.}~\bibnamefont{Yang}}, \bibinfo{journal}{J.
  Chem. Phys.} \textbf{\bibinfo{volume}{{109}}}, \bibinfo{pages}{10107}
  (\bibinfo{year}{1998}).

\bibitem[{\citenamefont{Levy and Perdew}(1985{\natexlab{b}})}]{LevPer-PRA-85}
\bibinfo{author}{\bibfnamefont{M.}~\bibnamefont{Levy}} \bibnamefont{and}
  \bibinfo{author}{\bibfnamefont{J.~P.} \bibnamefont{Perdew}},
  \bibinfo{journal}{Phys. Rev. A} \textbf{\bibinfo{volume}{32}},
  \bibinfo{pages}{2010} (\bibinfo{year}{1985}{\natexlab{b}}).

\bibitem[{\citenamefont{Liu and Burke}(2009)}]{LiuBur-JCP-09}
\bibinfo{author}{\bibfnamefont{Z.~F.} \bibnamefont{Liu}} \bibnamefont{and}
  \bibinfo{author}{\bibfnamefont{K.}~\bibnamefont{Burke}}, \bibinfo{journal}{J.
  Chem. Phys.} \textbf{\bibinfo{volume}{{131}}}, \bibinfo{pages}{124124}
  (\bibinfo{year}{2009}).

\bibitem[{\citenamefont{Gori-Giorgi
  et~al.}(2009{\natexlab{a}})\citenamefont{Gori-Giorgi, Seidl, and
  Vignale}}]{GorSeiVig-PRL-09}
\bibinfo{author}{\bibfnamefont{P.}~\bibnamefont{Gori-Giorgi}},
  \bibinfo{author}{\bibfnamefont{M.}~\bibnamefont{Seidl}}, \bibnamefont{and}
  \bibinfo{author}{\bibfnamefont{G.}~\bibnamefont{Vignale}},
  \bibinfo{journal}{Phys. Rev. Lett.} \textbf{\bibinfo{volume}{{103}}},
  \bibinfo{pages}{166402} (\bibinfo{year}{2009}{\natexlab{a}}).

\bibitem[{\citenamefont{Lieb}(1983)}]{Lie-IJQC-83}
\bibinfo{author}{\bibfnamefont{E.~H.} \bibnamefont{Lieb}},
  \bibinfo{journal}{Int. J. Quantum. Chem.} \textbf{\bibinfo{volume}{{24}}},
  \bibinfo{pages}{24} (\bibinfo{year}{1983}).

\bibitem[{\citenamefont{Seidl et~al.}(1999)\citenamefont{Seidl, Perdew, and
  Levy}}]{SeiPerLev-PRA-99}
\bibinfo{author}{\bibfnamefont{M.}~\bibnamefont{Seidl}},
  \bibinfo{author}{\bibfnamefont{J.~P.} \bibnamefont{Perdew}},
  \bibnamefont{and} \bibinfo{author}{\bibfnamefont{M.}~\bibnamefont{Levy}},
  \bibinfo{journal}{Phys. Rev. A} \textbf{\bibinfo{volume}{{59}}},
  \bibinfo{pages}{51} (\bibinfo{year}{1999}).

\bibitem[{\citenamefont{Seidl}(1999)}]{Sei-PRA-99}
\bibinfo{author}{\bibfnamefont{M.}~\bibnamefont{Seidl}},
  \bibinfo{journal}{Phys. Rev. A} \textbf{\bibinfo{volume}{{60}}},
  \bibinfo{pages}{4387} (\bibinfo{year}{1999}).

\bibitem[{\citenamefont{Seidl et~al.}(2000{\natexlab{a}})\citenamefont{Seidl,
  Perdew, and Kurth}}]{SeiPerKur-PRA-00}
\bibinfo{author}{\bibfnamefont{M.}~\bibnamefont{Seidl}},
  \bibinfo{author}{\bibfnamefont{J.~P.} \bibnamefont{Perdew}},
  \bibnamefont{and} \bibinfo{author}{\bibfnamefont{S.}~\bibnamefont{Kurth}},
  \bibinfo{journal}{Phys. Rev. A} \textbf{\bibinfo{volume}{{62}}},
  \bibinfo{pages}{012502} (\bibinfo{year}{2000}{\natexlab{a}}).

\bibitem[{\citenamefont{Seidl et~al.}(2007)\citenamefont{Seidl, Gori-Giorgi,
  and Savin}}]{SeiGorSav-PRA-07}
\bibinfo{author}{\bibfnamefont{M.}~\bibnamefont{Seidl}},
  \bibinfo{author}{\bibfnamefont{P.}~\bibnamefont{Gori-Giorgi}},
  \bibnamefont{and} \bibinfo{author}{\bibfnamefont{A.}~\bibnamefont{Savin}},
  \bibinfo{journal}{Phys. Rev. A} \textbf{\bibinfo{volume}{{75}}},
  \bibinfo{pages}{042511} (\bibinfo{year}{2007}).

\bibitem[{\citenamefont{Freund et~al.}(1984)\citenamefont{Freund, Huxtable, and
  Morgan}}]{FreHuxMor-PRA-84}
\bibinfo{author}{\bibfnamefont{D.~E.} \bibnamefont{Freund}},
  \bibinfo{author}{\bibfnamefont{B.~D.} \bibnamefont{Huxtable}},
  \bibnamefont{and} \bibinfo{author}{\bibfnamefont{J.~D.}
  \bibnamefont{Morgan}}, \bibinfo{journal}{Phys. Rev. A}
  \textbf{\bibinfo{volume}{{29}}}, \bibinfo{pages}{980} (\bibinfo{year}{1984}).

\bibitem[{\citenamefont{Gori-Giorgi and Savin}(2005)}]{GorSav-PRA-05}
\bibinfo{author}{\bibfnamefont{P.}~\bibnamefont{Gori-Giorgi}} \bibnamefont{and}
  \bibinfo{author}{\bibfnamefont{A.}~\bibnamefont{Savin}},
  \bibinfo{journal}{Phys. Rev. A} \textbf{\bibinfo{volume}{{71}}},
  \bibinfo{pages}{032513} (\bibinfo{year}{2005}).

\bibitem[{\citenamefont{Gori-Giorgi et~al.}(2008)\citenamefont{Gori-Giorgi,
  Seidl, and Savin}}]{GorSeiSav-PCCP-08}
\bibinfo{author}{\bibfnamefont{P.}~\bibnamefont{Gori-Giorgi}},
  \bibinfo{author}{\bibfnamefont{M.}~\bibnamefont{Seidl}}, \bibnamefont{and}
  \bibinfo{author}{\bibfnamefont{A.}~\bibnamefont{Savin}},
  \bibinfo{journal}{Phys. Chem. Chem. Phys.} \textbf{\bibinfo{volume}{{10}}},
  \bibinfo{pages}{3440} (\bibinfo{year}{2008}).

\bibitem[{\citenamefont{Levy}(1987)}]{Lev-INC-87}
\bibinfo{author}{\bibfnamefont{M.}~\bibnamefont{Levy}}, in
  \emph{\bibinfo{booktitle}{The single-Particle Density in Physics and
  Chemistry}}, edited by
  \bibinfo{editor}{\bibfnamefont{N.}~\bibnamefont{March}} \bibnamefont{and}
  \bibinfo{editor}{\bibfnamefont{B.}~\bibnamefont{Deb}}
  (\bibinfo{publisher}{Academic Press}, \bibinfo{address}{London},
  \bibinfo{year}{1987}).

\bibitem[{\citenamefont{Gori-Giorgi
  et~al.}(2009{\natexlab{b}})\citenamefont{Gori-Giorgi, Vignale, and
  Seidl}}]{GorVigSei-JCTC-09}
\bibinfo{author}{\bibfnamefont{P.}~\bibnamefont{Gori-Giorgi}},
  \bibinfo{author}{\bibfnamefont{G.}~\bibnamefont{Vignale}}, \bibnamefont{and}
  \bibinfo{author}{\bibfnamefont{M.}~\bibnamefont{Seidl}}, \bibinfo{journal}{J.
  Chem. Theory Comput.} \textbf{\bibinfo{volume}{{5}}}, \bibinfo{pages}{743}
  (\bibinfo{year}{2009}{\natexlab{b}}).

\bibitem[{\citenamefont{Attaccalite et~al.}(2002)\citenamefont{Attaccalite,
  Moroni, Gori-Giorgi, and Bachelet}}]{AttMorGorBac-PRL-02}
\bibinfo{author}{\bibfnamefont{C.}~\bibnamefont{Attaccalite}},
  \bibinfo{author}{\bibfnamefont{S.}~\bibnamefont{Moroni}},
  \bibinfo{author}{\bibfnamefont{P.}~\bibnamefont{Gori-Giorgi}},
  \bibnamefont{and} \bibinfo{author}{\bibfnamefont{G.~B.}
  \bibnamefont{Bachelet}}, \bibinfo{journal}{Phys. Rev. Lett.}
  \textbf{\bibinfo{volume}{88}}, \bibinfo{pages}{256601}
  (\bibinfo{year}{2002}).

\bibitem[{\citenamefont{Ceperley and Alder}(1980)}]{CepAld-PRL-80}
\bibinfo{author}{\bibfnamefont{D.~M.} \bibnamefont{Ceperley}} \bibnamefont{and}
  \bibinfo{author}{\bibfnamefont{B.~J.} \bibnamefont{Alder}},
  \bibinfo{journal}{Phys. Rev. Lett.} \textbf{\bibinfo{volume}{45}},
  \bibinfo{pages}{566} (\bibinfo{year}{1980}).

\bibitem[{\citenamefont{Vosko et~al.}(1980)\citenamefont{Vosko, Wilk, and
  Nusair}}]{VosWilNus-CJP-80}
\bibinfo{author}{\bibfnamefont{S.~J.} \bibnamefont{Vosko}},
  \bibinfo{author}{\bibfnamefont{L.}~\bibnamefont{Wilk}}, \bibnamefont{and}
  \bibinfo{author}{\bibfnamefont{M.}~\bibnamefont{Nusair}},
  \bibinfo{journal}{Can. J. Phys.} \textbf{\bibinfo{volume}{{58}}},
  \bibinfo{pages}{1200} (\bibinfo{year}{1980}).

\bibitem[{\citenamefont{Perdew and Wang}(1992)}]{PerWan-PRB-92}
\bibinfo{author}{\bibfnamefont{J.~P.} \bibnamefont{Perdew}} \bibnamefont{and}
  \bibinfo{author}{\bibfnamefont{Y.}~\bibnamefont{Wang}},
  \bibinfo{journal}{Phys. Rev. B} \textbf{\bibinfo{volume}{45}},
  \bibinfo{pages}{13244} (\bibinfo{year}{1992}).

\bibitem[{\citenamefont{Sun et~al.}(2010)\citenamefont{Sun, Perdew, and
  Seidl}}]{SunPerSei-PRB-10}
\bibinfo{author}{\bibfnamefont{J.}~\bibnamefont{Sun}},
  \bibinfo{author}{\bibfnamefont{J.~P.} \bibnamefont{Perdew}},
  \bibnamefont{and} \bibinfo{author}{\bibfnamefont{M.}~\bibnamefont{Seidl}},
  \bibinfo{journal}{Phys. Rev. B} \textbf{\bibinfo{volume}{{81}}},
  \bibinfo{pages}{085123} (\bibinfo{year}{2010}).

\bibitem[{\citenamefont{Casula et~al.}(2006)\citenamefont{Casula, Sorella, and
  Senatore}}]{CasSorSen-PRB-06}
\bibinfo{author}{\bibfnamefont{M.}~\bibnamefont{Casula}},
  \bibinfo{author}{\bibfnamefont{S.}~\bibnamefont{Sorella}}, \bibnamefont{and}
  \bibinfo{author}{\bibfnamefont{G.}~\bibnamefont{Senatore}},
  \bibinfo{journal}{Phys. Rev. B} \textbf{\bibinfo{volume}{{74}}},
  \bibinfo{pages}{245427} (\bibinfo{year}{2006}).

\bibitem[{\citenamefont{{von Barth} and Hedin}(1972)}]{BarHed-JPC-72}
\bibinfo{author}{\bibfnamefont{U.}~\bibnamefont{{von Barth}}} \bibnamefont{and}
  \bibinfo{author}{\bibfnamefont{L.}~\bibnamefont{Hedin}}, \bibinfo{journal}{J.
  Phys. C} \textbf{\bibinfo{volume}{{5}}}, \bibinfo{pages}{1629}
  (\bibinfo{year}{1972}).

\bibitem[{\citenamefont{van Leeuwen and Baerends}(1994)}]{vanBae-PRA-94}
\bibinfo{author}{\bibfnamefont{R.}~\bibnamefont{van Leeuwen}} \bibnamefont{and}
  \bibinfo{author}{\bibfnamefont{E.~J.} \bibnamefont{Baerends}},
  \bibinfo{journal}{Phys. Rev. A} \textbf{\bibinfo{volume}{{49}}},
  \bibinfo{pages}{2421} (\bibinfo{year}{1994}).

\bibitem[{\citenamefont{Zhao et~al.}(1994)\citenamefont{Zhao, Morrison, and
  Parr}}]{ZhaMorPar-PRA-94}
\bibinfo{author}{\bibfnamefont{Q.}~\bibnamefont{Zhao}},
  \bibinfo{author}{\bibfnamefont{R.~C.} \bibnamefont{Morrison}},
  \bibnamefont{and} \bibinfo{author}{\bibfnamefont{R.~G.} \bibnamefont{Parr}},
  \bibinfo{journal}{Phys. Rev. A} \textbf{\bibinfo{volume}{{50}}},
  \bibinfo{pages}{2138} (\bibinfo{year}{1994}).

\bibitem[{\citenamefont{Colonna and Savin}(1999)}]{ColSav-JCP-99}
\bibinfo{author}{\bibfnamefont{F.}~\bibnamefont{Colonna}} \bibnamefont{and}
  \bibinfo{author}{\bibfnamefont{A.}~\bibnamefont{Savin}}, \bibinfo{journal}{J.
  Chem. Phys.} \textbf{\bibinfo{volume}{{110}}}, \bibinfo{pages}{2828}
  (\bibinfo{year}{1999}).

\bibitem[{\citenamefont{Buttazzo et~al.}(2010)\citenamefont{Buttazzo, {De
  Pascale}, and Gori-Giorgi}}]{ButDepGor-PRA-10}
\bibinfo{author}{\bibfnamefont{G.}~\bibnamefont{Buttazzo}},
  \bibinfo{author}{\bibfnamefont{L.}~\bibnamefont{{De Pascale}}},
  \bibnamefont{and}
  \bibinfo{author}{\bibfnamefont{P.}~\bibnamefont{Gori-Giorgi}},
  \bibinfo{journal}{in preparation} \textbf{\bibinfo{volume}{{}}}
  (\bibinfo{year}{2010}).

\bibitem[{\citenamefont{Reimann and Manninen}(2002)}]{ReiMan-RMP-02}
\bibinfo{author}{\bibfnamefont{S.~M.} \bibnamefont{Reimann}} \bibnamefont{and}
  \bibinfo{author}{\bibfnamefont{M.}~\bibnamefont{Manninen}},
  \bibinfo{journal}{Rev. Mod. Phys.} \textbf{\bibinfo{volume}{{74}}},
  \bibinfo{pages}{1283} (\bibinfo{year}{2002}).

\bibitem[{\citenamefont{Jiang et~al.}(2003)\citenamefont{Jiang, Baranger, and
  Yang}}]{JiaBarYan-PRB-03}
\bibinfo{author}{\bibfnamefont{H.}~\bibnamefont{Jiang}},
  \bibinfo{author}{\bibfnamefont{H.~U.} \bibnamefont{Baranger}},
  \bibnamefont{and} \bibinfo{author}{\bibfnamefont{W.}~\bibnamefont{Yang}},
  \bibinfo{journal}{Phys. Rev. B} \textbf{\bibinfo{volume}{68}},
  \bibinfo{pages}{165337} (\bibinfo{year}{2003}).

\bibitem[{\citenamefont{Jiang et~al.}(2004)\citenamefont{Jiang, Ullmo, Yang,
  and Baranger}}]{JiaUllYanBar-PRB-04}
\bibinfo{author}{\bibfnamefont{H.}~\bibnamefont{Jiang}},
  \bibinfo{author}{\bibfnamefont{D.}~\bibnamefont{Ullmo}},
  \bibinfo{author}{\bibfnamefont{W.}~\bibnamefont{Yang}}, \bibnamefont{and}
  \bibinfo{author}{\bibfnamefont{H.~U.} \bibnamefont{Baranger}},
  \bibinfo{journal}{Phys. Rev. B} \textbf{\bibinfo{volume}{69}},
  \bibinfo{pages}{235326} (\bibinfo{year}{2004}).

\bibitem[{\citenamefont{R\"as\"anen et~al.}(2004)\citenamefont{R\"as\"anen,
  Harju, Puska, and Nieminen}}]{RasHarPusNie-PRB-04}
\bibinfo{author}{\bibfnamefont{E.}~\bibnamefont{R\"as\"anen}},
  \bibinfo{author}{\bibfnamefont{A.}~\bibnamefont{Harju}},
  \bibinfo{author}{\bibfnamefont{M.~J.} \bibnamefont{Puska}}, \bibnamefont{and}
  \bibinfo{author}{\bibfnamefont{R.~M.} \bibnamefont{Nieminen}},
  \bibinfo{journal}{Phys. Rev. B} \textbf{\bibinfo{volume}{69}},
  \bibinfo{pages}{165309} (\bibinfo{year}{2004}).

\bibitem[{\citenamefont{Pittalis et~al.}(2009)\citenamefont{Pittalis,
  R\"as\"anen, Proetto, and Gross}}]{PitRasProGro-PRB-09}
\bibinfo{author}{\bibfnamefont{S.}~\bibnamefont{Pittalis}},
  \bibinfo{author}{\bibfnamefont{E.}~\bibnamefont{R\"as\"anen}},
  \bibinfo{author}{\bibfnamefont{C.~R.} \bibnamefont{Proetto}},
  \bibnamefont{and} \bibinfo{author}{\bibfnamefont{E.~K.~U.}
  \bibnamefont{Gross}}, \bibinfo{journal}{Phys. Rev. B}
  \textbf{\bibinfo{volume}{79}}, \bibinfo{pages}{085316}
  (\bibinfo{year}{2009}).

\bibitem[{\citenamefont{Rontani et~al.}(2006)\citenamefont{Rontani, Cavazzoni,
  Bellucci, and Goldoni}}]{RonCavBelGol-JCP-06}
\bibinfo{author}{\bibfnamefont{M.}~\bibnamefont{Rontani}},
  \bibinfo{author}{\bibfnamefont{C.}~\bibnamefont{Cavazzoni}},
  \bibinfo{author}{\bibfnamefont{D.}~\bibnamefont{Bellucci}}, \bibnamefont{and}
  \bibinfo{author}{\bibfnamefont{G.}~\bibnamefont{Goldoni}},
  \bibinfo{journal}{J. Chem. Phys.} \textbf{\bibinfo{volume}{{124}}},
  \bibinfo{pages}{124102} (\bibinfo{year}{2006}).

\bibitem[{\citenamefont{Blundell and Joshi}(2010)}]{BluJos-PRB-10}
\bibinfo{author}{\bibfnamefont{S.~A.} \bibnamefont{Blundell}} \bibnamefont{and}
  \bibinfo{author}{\bibfnamefont{K.}~\bibnamefont{Joshi}},
  \bibinfo{journal}{Phys. Rev. B} \textbf{\bibinfo{volume}{{81}}},
  \bibinfo{pages}{115323} (\bibinfo{year}{2010}).

\bibitem[{\citenamefont{Ghosal et~al.}(2006)\citenamefont{Ghosal, Guclu,
  Umrigar, Ullmo, and Baranger}}]{GhoGucUmrUllBar-NP-06}
\bibinfo{author}{\bibfnamefont{A.}~\bibnamefont{Ghosal}},
  \bibinfo{author}{\bibfnamefont{A.~D.} \bibnamefont{Guclu}},
  \bibinfo{author}{\bibfnamefont{C.~J.} \bibnamefont{Umrigar}},
  \bibinfo{author}{\bibfnamefont{D.}~\bibnamefont{Ullmo}}, \bibnamefont{and}
  \bibinfo{author}{\bibfnamefont{H.~U.} \bibnamefont{Baranger}},
  \bibinfo{journal}{Nature Phys.} \textbf{\bibinfo{volume}{2}},
  \bibinfo{pages}{336} (\bibinfo{year}{2006}).

\bibitem[{\citenamefont{Zeng et~al.}(2009)\citenamefont{Zeng, Geist, Ruan,
  Umrigar, and Chou}}]{ZenGeiRuaUmrCho-PRB-09}
\bibinfo{author}{\bibfnamefont{L.}~\bibnamefont{Zeng}},
  \bibinfo{author}{\bibfnamefont{W.}~\bibnamefont{Geist}},
  \bibinfo{author}{\bibfnamefont{W.~Y.} \bibnamefont{Ruan}},
  \bibinfo{author}{\bibfnamefont{C.~J.} \bibnamefont{Umrigar}},
  \bibnamefont{and} \bibinfo{author}{\bibfnamefont{M.~Y.} \bibnamefont{Chou}},
  \bibinfo{journal}{Phys. Rev. B} \textbf{\bibinfo{volume}{79}},
  \bibinfo{pages}{235334} (\bibinfo{year}{2009}).

\bibitem[{\citenamefont{Guclu et~al.}(2008)\citenamefont{Guclu, Ghosal,
  Umrigar, and Baranger}}]{GucGhoUmrBar-PRB-08}
\bibinfo{author}{\bibfnamefont{A.~D.} \bibnamefont{Guclu}},
  \bibinfo{author}{\bibfnamefont{A.}~\bibnamefont{Ghosal}},
  \bibinfo{author}{\bibfnamefont{C.~J.} \bibnamefont{Umrigar}},
  \bibnamefont{and} \bibinfo{author}{\bibfnamefont{H.~U.}
  \bibnamefont{Baranger}}, \bibinfo{journal}{Phys. Rev. B}
  \textbf{\bibinfo{volume}{77}}, \bibinfo{pages}{041301}
  (\bibinfo{year}{2008}).

\bibitem[{\citenamefont{Yannouleas and Landman}(2007)}]{YanLan-RPP-07}
\bibinfo{author}{\bibfnamefont{C.}~\bibnamefont{Yannouleas}} \bibnamefont{and}
  \bibinfo{author}{\bibfnamefont{U.}~\bibnamefont{Landman}},
  \bibinfo{journal}{Rep. Prog. Phys.} \textbf{\bibinfo{volume}{{70}}},
  \bibinfo{pages}{2067} (\bibinfo{year}{2007}).

\bibitem[{\citenamefont{Ziesche et~al.}(2000)\citenamefont{Ziesche, Tao, Seidl,
  and Perdew}}]{ZieTaoSeiPer-IJQC-00}
\bibinfo{author}{\bibfnamefont{P.}~\bibnamefont{Ziesche}},
  \bibinfo{author}{\bibfnamefont{J.}~\bibnamefont{Tao}},
  \bibinfo{author}{\bibfnamefont{M.}~\bibnamefont{Seidl}}, \bibnamefont{and}
  \bibinfo{author}{\bibfnamefont{J.~P.} \bibnamefont{Perdew}},
  \bibinfo{journal}{Int. J. Quantum Chem.} \textbf{\bibinfo{volume}{{77}}},
  \bibinfo{pages}{819} (\bibinfo{year}{2000}).

\bibitem[{\citenamefont{Taut}(1994)}]{Tau-PAMG-94}
\bibinfo{author}{\bibfnamefont{M.}~\bibnamefont{Taut}}, \bibinfo{journal}{Phys.
  A: Math. Gen.} \textbf{\bibinfo{volume}{{27}}}, \bibinfo{pages}{1045}
  (\bibinfo{year}{1994}).

\bibitem[{\citenamefont{Drummond and Needs}(2009)}]{DruNee-PRL-09}
\bibinfo{author}{\bibfnamefont{N.~D.} \bibnamefont{Drummond}} \bibnamefont{and}
  \bibinfo{author}{\bibfnamefont{R.~J.} \bibnamefont{Needs}},
  \bibinfo{journal}{Phys. Rev. Lett.} \textbf{\bibinfo{volume}{{102}}},
  \bibinfo{pages}{126402} (\bibinfo{year}{2009}).

\bibitem[{\citenamefont{Carr}(1961)}]{Car-PR-61}
\bibinfo{author}{\bibfnamefont{W.~J.} \bibnamefont{Carr}},
  \bibinfo{journal}{Phys. Rev.} \textbf{\bibinfo{volume}{{122}}},
  \bibinfo{pages}{1437} (\bibinfo{year}{1961}).

\bibitem[{\citenamefont{Teale et~al.}(2010)\citenamefont{Teale, Coriani, and
  Helgaker}}]{TeaCorHel-JCP-10}
\bibinfo{author}{\bibfnamefont{A.~M.} \bibnamefont{Teale}},
  \bibinfo{author}{\bibfnamefont{S.}~\bibnamefont{Coriani}}, \bibnamefont{and}
  \bibinfo{author}{\bibfnamefont{T.}~\bibnamefont{Helgaker}},
  \bibinfo{journal}{J. Chem. Phys.} \textbf{\bibinfo{volume}{{132}}},
  \bibinfo{pages}{164115} (\bibinfo{year}{2010}).

\bibitem[{\citenamefont{Seidl et~al.}(2000{\natexlab{b}})\citenamefont{Seidl,
  Perdew, and Kurth}}]{SeiPerKur-PRL-00}
\bibinfo{author}{\bibfnamefont{M.}~\bibnamefont{Seidl}},
  \bibinfo{author}{\bibfnamefont{J.~P.} \bibnamefont{Perdew}},
  \bibnamefont{and} \bibinfo{author}{\bibfnamefont{S.}~\bibnamefont{Kurth}},
  \bibinfo{journal}{Phys. Rev. Lett.} \textbf{\bibinfo{volume}{{84}}},
  \bibinfo{pages}{5070} (\bibinfo{year}{2000}{\natexlab{b}}).

\bibitem[{\citenamefont{G\"{o}rling and Levy}(1994)}]{GorLev-PRA-94}
\bibinfo{author}{\bibfnamefont{A.}~\bibnamefont{G\"{o}rling}} \bibnamefont{and}
  \bibinfo{author}{\bibfnamefont{M.}~\bibnamefont{Levy}},
  \bibinfo{journal}{Phys. Rev. A} \textbf{\bibinfo{volume}{50}},
  \bibinfo{pages}{196} (\bibinfo{year}{1994}).

\bibitem[{\citenamefont{Gori-Giorgi and Savin}(2008)}]{GorSav-JPCS-08}
\bibinfo{author}{\bibfnamefont{P.}~\bibnamefont{Gori-Giorgi}} \bibnamefont{and}
  \bibinfo{author}{\bibfnamefont{A.}~\bibnamefont{Savin}}, \bibinfo{journal}{J.
  Phys.: Conf. Ser.} \textbf{\bibinfo{volume}{{117}}}, \bibinfo{pages}{012017}
  (\bibinfo{year}{2008}).

\bibitem[{\citenamefont{Savin}(2009)}]{Sav-CP-09}
\bibinfo{author}{\bibfnamefont{A.}~\bibnamefont{Savin}},
  \bibinfo{journal}{Chem. Phys.} \textbf{\bibinfo{volume}{{356}}},
  \bibinfo{pages}{91} (\bibinfo{year}{2009}).

\bibitem[{\citenamefont{Becke and Roussel}(1989)}]{BecRou-PRA-89}
\bibinfo{author}{\bibfnamefont{A.~D.} \bibnamefont{Becke}} \bibnamefont{and}
  \bibinfo{author}{\bibfnamefont{M.~R.} \bibnamefont{Roussel}},
  \bibinfo{journal}{Phys. Rev. A} \textbf{\bibinfo{volume}{{39}}},
  \bibinfo{pages}{3761} (\bibinfo{year}{1989}).

\bibitem[{\citenamefont{Becke}(2003)}]{Bec-JCP-03}
\bibinfo{author}{\bibfnamefont{A.~D.} \bibnamefont{Becke}},
  \bibinfo{journal}{J. Chem. Phys.} \textbf{\bibinfo{volume}{{119}}},
  \bibinfo{pages}{2972} (\bibinfo{year}{2003}).

\bibitem[{\citenamefont{R\"as\"anen et~al.}(2010)\citenamefont{R\"as\"anen,
  Seidl, and Gori-Giorgi}}]{RasSeiGor-XXX-10}
\bibinfo{author}{\bibfnamefont{E.}~\bibnamefont{R\"as\"anen}},
  \bibinfo{author}{\bibfnamefont{M.}~\bibnamefont{Seidl}}, \bibnamefont{and}
  \bibinfo{author}{\bibfnamefont{P.}~\bibnamefont{Gori-Giorgi}},
  \bibinfo{journal}{in preparation} \textbf{\bibinfo{volume}{{}}}
  (\bibinfo{year}{2010}).

\end{thebibliography}
\end{document}